\documentclass[%
 reprint,
 amsmath,amssymb,
 aps,
showkeys
]{revtex4-1}

\usepackage{graphicx}% Include figure files
\usepackage{dcolumn}% Align table columns on decimal point
\usepackage{bm}% bold math
\usepackage{xcolor}

\begin{document}

%\preprint{APS/123-QED}

\title{On the Dissipation Rate of Temperature Fluctuations in Stably Stratified Flows}% Force line breaks with \\

\author{Sukanta Basu}
\email{sukanta.basu@gmail.com}
\affiliation{Faculty of Civil Engineering and Geosciences, Delft University of Technology, Delft, the Netherlands
}
\author{Adam W. DeMarco}
\email{awdemarc@ncsu.edu}
\affiliation{United States Air Force, USA }%

\author{Ping He}%
\email{drpinghe@umich.edu}
\affiliation{Department of Aerospace Engineering, University of Michigan, Ann Arbor, USA}

\date{\today}% It is always \today, today,

\begin{abstract}
In this study, we explore several integral and outer length scales of turbulence which can be formulated by using the dissipation of temperature fluctuations ($\chi$) and other relevant variables. Our analyses directly lead to simple yet non-trivial parameterizations for both spatially-averaged $\overline{\chi}$ and the structure parameter of temperature ($C_T^2$). For our purposes, we make use of high-fidelity data from direct numerical simulations of stratified channel flows. 
\end{abstract}

%Use showkeys class option if keyword
\maketitle

\section{Introduction}
\label{intro}

The molecular dissipation of temperature fluctuations ($\chi$) is an important variable for characterizing turbulent mixing in various environmental flows. It is frequently used in micrometeorology \cite[e.g.,][]{wyngaard71a} and atmospheric optics \cite[e.g.,][]{muschinski15}. Furthermore, any higher-order closure model requires solving a prognostic equation or a diagnostic parameterization for ensemble-averaged $\overline{\chi}$ \cite[refer to][]{mellor82,kantha94,cheng02,wyngaard10}.  

Over the years, a number of studies focused on the correlation between turbulent kinetic energy dissipation rate ($\varepsilon$) and $\chi$ \cite[e.g.,][]{antonia75,antonia80,zhou00,antonia01,hao08,abe09}. In addition, some papers reported on the probability density function, spatio-temporal intermittency and anomalous scaling of $\chi$ \cite[e.g.,][]{antonia75,schmitt96,sreenivasan97}. Often, $\chi$ has been found to be more intermittent (commonly quantified by the multifractal scaling exponents) and non-Gaussian than $\varepsilon$ \citep[e.g.,][]{prasad88,sreenivasan97}. 

Most of these previous studies primarily focused on the instantaneous, localized traits of the dissipation fields. Instead, we are  interested to better quantify their spatially averaged characteristics. Towards this goal, we first investigate the statistical properties of several length scales which can be formulated based on $\overline{\chi}$ and other relevant variables. Based on these findings, we then derive simple parameterizations for $\overline{\chi}$ and temperature structure parameter ($C_T^2$). For all the analyses, we utilize a direct numerical simulation (DNS) database of stably stratified flows which is discussed in the following section.\\

\section{Direct Numerical Simulation}
\label{DNS}

Recently, for the parameterization of optical turbulence, He and Basu~\cite{he16c} created a DNS database using a massively parallel DNS code, called HERCULES \citep{he16a}. The DNS runs were conducted by solving the normalized Navier$-$Stokes and temperature equations, as shown in Eqs.~\ref{eqn_mass} to \ref{eqn_energy} (using Einstein's summation notation): 

\begin{widetext}
\begin{equation}
\label{eqn_mass}
\frac{\partial u_{n, i}}{\partial x_{n, i}} =0,
\end{equation}

\begin{equation}
\label{eqn_momentum}
\frac{\partial u_{n, i}}{\partial t_n}+\frac{\partial u_{n, i} u_{n, j}}{\partial x_{n, j}}=-\frac{\partial p_n}{\partial x_{n, i}} +\frac{1}{Re_b} \frac{\partial }{\partial x_{n, j}} \left(\frac{\partial u_{n, i}}{\partial x_{n, j}}\right) + \mathrm{\Delta}P {\delta}_{i1}+Ri_b {\theta_n} {\delta}_{i3}, 
\end{equation}

\begin{equation}
\label{eqn_energy}
\frac{\partial \theta_n}{\partial t_n}+\frac{\partial \theta_n u_{n, i}}{\partial x_{n, i}}=\frac{1}{Re_b Pr} \frac{\partial }{\partial x_{n, i}} \left(\frac{\partial \theta_n}{\partial x_{n, i}}\right),
\end{equation}
\end{widetext}
where $u_n$ and $x_n$ are the normalized velocity and coordinate vectors, respectively, with the subscript $i$ denoting the $i$\textsuperscript{th} vector component; $t_n$ is the normalized time; $p_n$ is the normalized pressure; $\mathrm{\Delta}P$ is the streamwise pressure gradient to drive the flow; and $\theta_n$ is the normalized potential temperature. The bulk Richardson number is denoted by:
\begin{equation}
Ri_b = \frac{\left(\Theta_{top}-\Theta_{bot}\right)g h}{U_b^2 \Theta_{top}},     
\end{equation}
where, $g$ denotes the gravitational acceleration, and $\Theta_{top}$ and $\Theta_{bot}$ represent potential temperature at the top and the bottom of the channel, respectively. $Pr = \nu/k = 0.7$ is the Prandtl number with $k$ being the thermal diffusivity, and $Re_b = \frac{U_b h}{\nu}$ is the bulk Reynolds number with $h$, $U_b$, and $\nu$ being the channel height, the bulk (averaged) velocity in the channel, and the kinematic viscosity, respectively. The bulk Reynolds number was fixed at 20,000 for all the simulations. 

The computational domain size for all the DNS runs was $L_x \times L_y \times L_z = 18 h \times 10 h \times h$. The domain was discretized by $2304 \times 2048 \times 288$ grid points in streamwise, spanwise, and wall-normal directions, respectively. A total of five simulations were performed with gradual increase in the temperature difference between the top and bottom walls (effectively by increasing $Ri_b$) to mimic the nighttime cooling of the land-surface. The normalized cooling rates ($CR$), $\partial Ri_b/ \partial T_n$, ranged from $1\times10^{-3}$ to $5\times10^{-3}$; where, $T_n$ is a non-dimensional time ($=tU_b/h$). 

All the simulations used fully developed neutrally stratified flows ($Ri_b = 0$) as initial conditions and evolved for up to $T_n = 100$. The simulation results were output every 10 non-dimensional time. To avoid spin-up issues, in the present study, we only use data for the last five output files (i.e., $60 \le T_n \le 100$). Furthermore, we only consider data from the region $0.1 h\le z \le 0.5 h$ to discard any blocking effect of the surface or avoid any laminarization in the upper part of the open channel. Vertical profiles and some basic statistics from these simulations have been documented in Appendix~3. 

The mean dissipation of turbulent kinetic energy and temperature fluctuations are computed as follows: 
\begin{subequations}
\begin{equation}
\overline{\varepsilon} = \nu \overline{\left(\frac{\partial u_i'}{\partial x_j} \frac{\partial u_i'}{\partial x_j}\right)},
\end{equation}
\begin{equation}
\overline{\chi} = 2k \overline{\left(\frac{\partial \theta'}{\partial x_j} \frac{\partial \theta'}{\partial x_j} \right)}.
\end{equation}
\end{subequations}
In the above equations, and in the rest of the paper, the ``overbar'' notation is used to denote mean quantities. Horizontal (planar) averaging operation is performed for all the cases. The ``prime'' symbol is used to represent the fluctuation of a variable with respect to its planar averaged value. 

In a recent paper, Basu et al.~\cite{basu20} utilized this DNS database to derive parameterizations for $\overline{\varepsilon}$. In the present work, the focus is placed on $\overline{\chi}$. 

\section{Integral Length Scales}

From the DNS-generated data, we first calculate two different integral length scales as follows: 
\begin{subequations}
\begin{equation}
    \mathcal{L} \equiv \frac{\overline{e}^{3/2}}{\overline{\varepsilon}},
\end{equation}
\begin{equation}
    \mathcal{L_\theta} \equiv \frac{\overline{e}^{1/2}\sigma_\theta^2}{\overline{\chi}},
\end{equation}
\label{EqILS}
\end{subequations}

\noindent where, $\overline{e}$ and $\sigma_\theta^2$ denote turbulent kinetic energy (TKE) and the variance of temperature, respectively. 

Based on the original ideas of Taylor \cite{taylor35}, both \cite{tennekes72}  and \cite{pope00} provided a heuristic derivation of $\mathcal{L}$. Given TKE ($\overline{e}$) and mean energy dissipation rate ($\overline{\varepsilon}$), an associated integral time scale can be approximated as $\overline{e}/\overline{\varepsilon}$. One can further assume $\sqrt{\overline{e}}$ to be the corresponding velocity scale. Thus, an integral length scale can be approximated as $\overline{e}^{3/2}/\overline{\varepsilon}$. Using dimensional arguments, an analogous length scale $\mathcal{L_\theta}$ can be formulated based on temperature fluctuations \citep{xu00,abe11}. 

\begin{figure*}[ht]
\centering
  \includegraphics[width=2.3in]{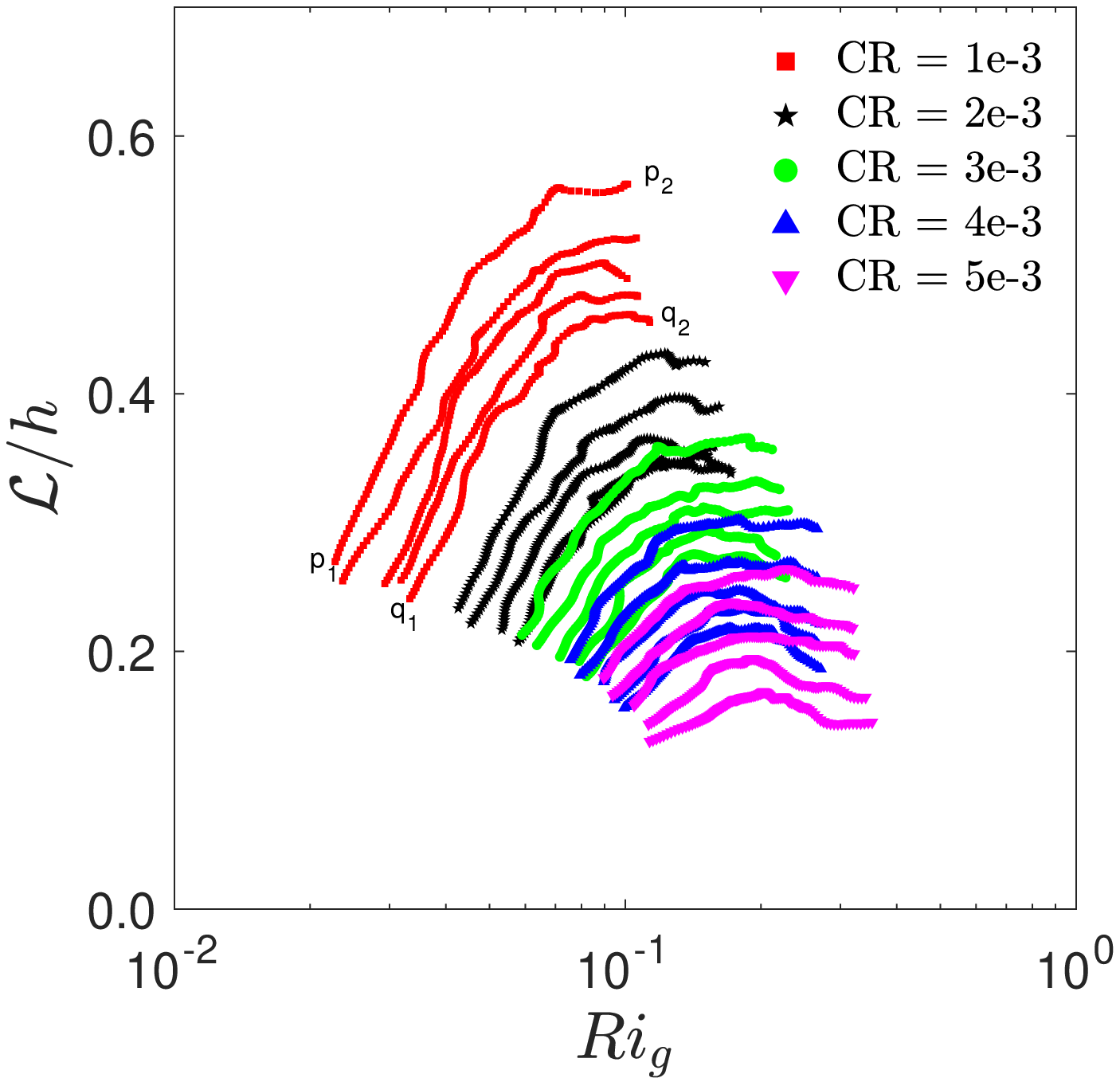}
  \hspace{0.4in}
  \includegraphics[width=2.3in]{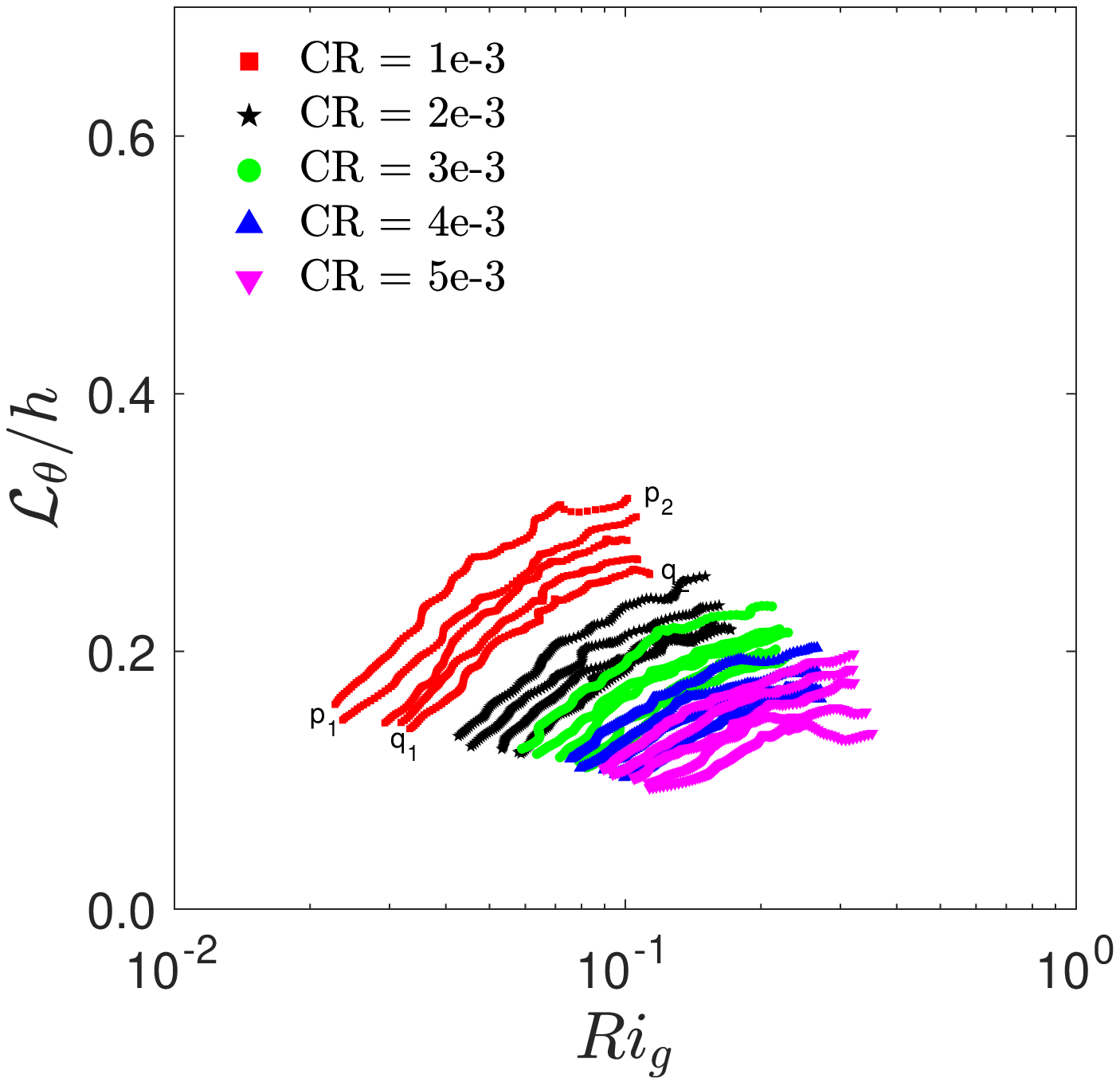}\\
  \includegraphics[width=2.3in]{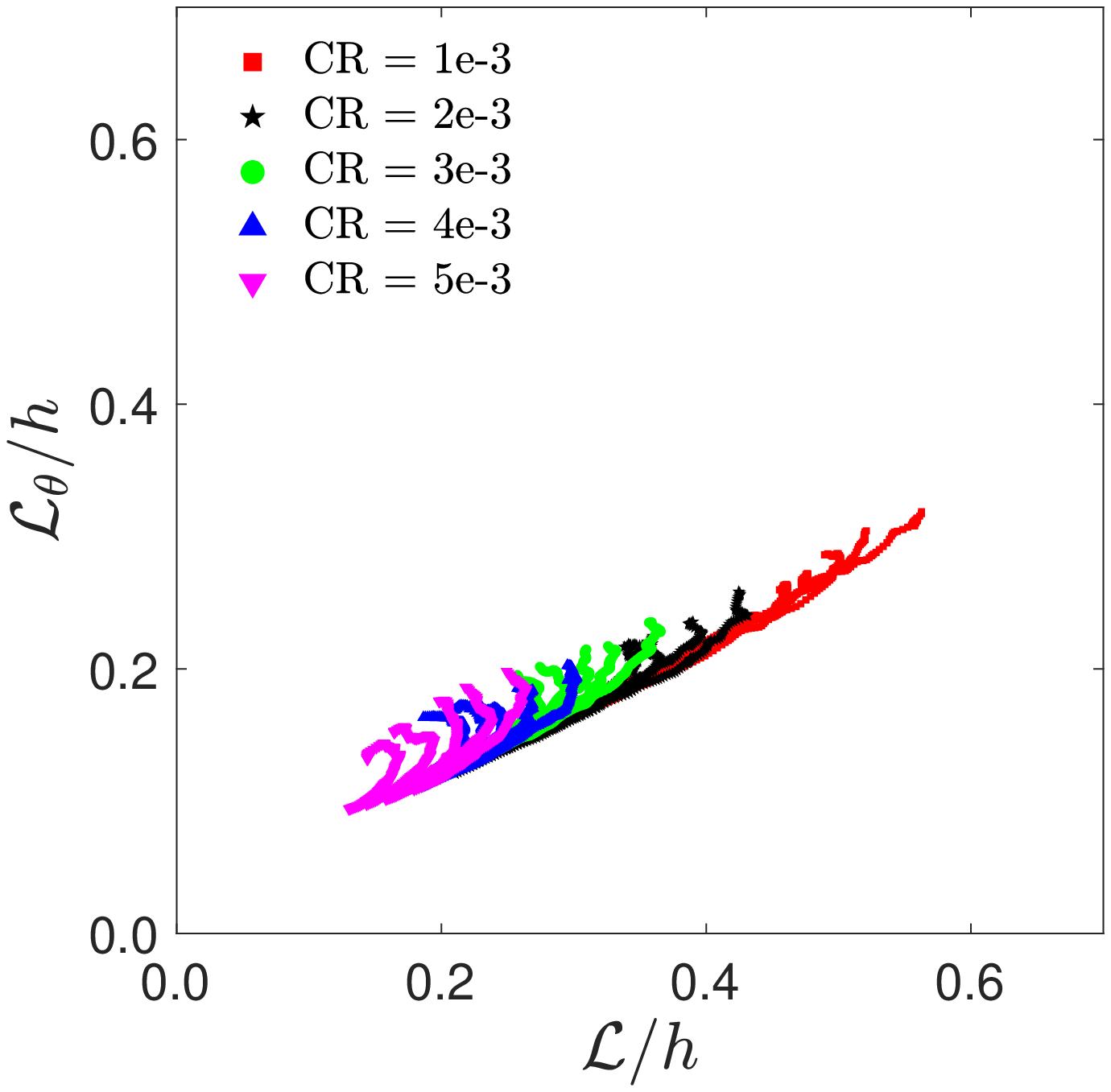}
  \hspace{0.4in}
  \includegraphics[width=2.3in]{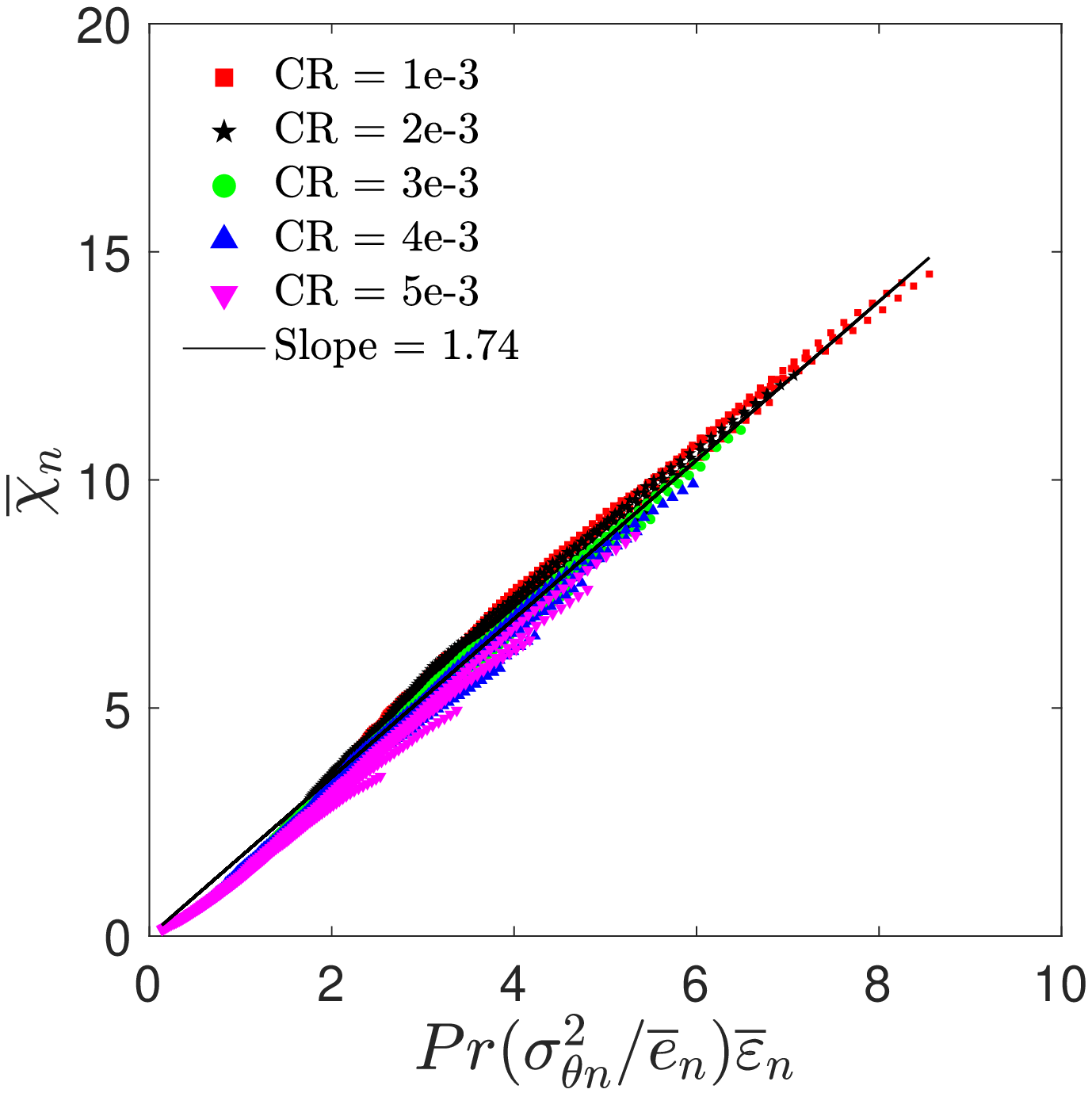}
\caption{Top panel: integral length scales as functions of gradient Richardson number. Both the length scales are normalized by the height of the open channel ($h$). Bottom-left panel: scatter plot of $\mathcal{L}$ vs. $\mathcal{L}_\theta$. Bottom-right panel: normalized $\overline{\chi}$ as a function of normalized $\overline{\varepsilon}$, $\overline{e}$, and $\sigma_\theta^2$. Please refer to Eq.~\ref{EDRCHI}. Simulated data from five different DNS runs are represented by different colored symbols in these plots. In the legends, $CR$ represents normalized cooling rates.}
\label{fig1}      
\end{figure*}

In the top-panel of Fig.~\ref{fig1}, normalized values of $\mathcal{L}$ and $\mathcal{L}_\theta$ are plotted against the gradient Richardson number ($Ri_g = N^2/S^2$); where, $N$ is the Brunt-V\"{a}is\"{a}la frequency and $S$ is the magnitude of wind shear. In these plots, we have marked four specific points based on the data from DNS run with imposed cooling rate of 10$^{-3}$ to better understand the effects of height and stability on the integral length scales. The points $p_1$ and $p_2$ represent data from $z/h = 0.1$ and $z/h = 0.5$, respectively at non-dimensional time ($T_n$) of 60. Similarly, $q_1$ and $q_2$ are associated with data from $z/h = 0.1$ and $z/h = 0.5$, respectively at non-dimensional time ($T_n$) of 100. 

Physically, one would expect the integral scales to be increasing with height as long as the eddies feel the presence of the surface (near-neutral or weakly stable condition). For very stable conditions, the eddies no longer feel the presence of the surface. In the atmospheric boundary layer literature, such a situation is known as the z-less condition \citep{wyngaard73,grisogono10}. Under the influence of strong stability, the integral length scales become more-or-less independent of the height above the surface.

From Fig.~\ref{fig1}, it is clear that the integral length scales increase with height and they slowly decrease with time in all the simulations due to the increasing stability effects. Simulations with higher cooling rates have smaller integral length scales. Some of these runs (e.g., $CR = 5\times10^{-3}$) exhibit z-less behavior due to strong stability effects.

Given the similar trends of normalized $\mathcal{L}$ and $\mathcal{L}_\theta$, they are plotted against each other in the bottom-left panel of Fig.~\ref{fig1}. There is (approximately) a linear relationship between these length scales. If $\mathcal{L} \propto \mathcal{L_\theta}$, it is straightforward to derive from Eqs.~\ref{EqILS}: 
\begin{equation}
    \frac{\overline{\chi}}{\overline{\varepsilon}} \propto \frac{\sigma_\theta^2}{\overline{e}}.
    \label{EDRCHI}
\end{equation}

This relationship was first reported by B\'{e}guier et al.~\cite{beguier78} for shear flow turbulence. In a follow-up study, Elghobashi and Launder~\cite{elghobashi83} hypothesized that the similarity of the generation processes of TKE and scalar variance is at the root of this intriguing relationship. In contrast to shear flows, they did not find Eq.~\ref{EDRCHI} to hold for thermal mixed layer. 

In the bottom-right panel of Fig.~\ref{fig1}, we demonstrate the approximate validity of Eq.~\ref{EDRCHI}. Linear least-square regression with bootstrapping \citep{efron82,mooney93} is used to estimate the slope of the fitted line. Given that the collapse of the data points is quite reasonable, the relationship $\overline{\chi} = 1.74 \overline{\varepsilon} \frac{\sigma_\theta^2}{\overline{e}}$ might be useful for practical applications. 

Please note that the appearance of the Prandtl number ($Pr$) in Fig.~\ref{fig1} (bottom-right panel) is due to the normalization of variables in DNS; Appendix~2 provides further details. Throughout the paper, the subscript ``$n$'' is used to denote a normalized variable. 

\section{Outer Length Scales}

Both shear and buoyancy prefer to deform larger eddies compared to smaller ones \citep{itsweire93,smyth00,chung12,mater13}. Turbulent eddies are not affected by shear and buoyancy if they are smaller than the outer length scales (OLSs). Ozmidov ($L_{OZ}$) and Corrsin ($L_C$) length scales are the most commonly used OLSs in the literature. They are defined as \citep{corrsin58,dougherty61,ozmidov65}: 
\begin{subequations}
\begin{equation}
    L_{OZ} \equiv \left(\frac{\overline{\varepsilon}}{N^3}\right)^{1/2},
\end{equation}
\begin{equation}
    L_C \equiv \left(\frac{\overline{\varepsilon}}{S^3}\right)^{1/2}.
    \label{LC}
\end{equation}
\label{OLS}
\end{subequations}

\noindent Eddies which are smaller than $L_{OZ}$ are not affected by buoyancy; similarly, shear does not influence the eddies of size less than $L_{C}$. In other words, the eddies can be assumed to be isotropic if they are smaller than both $L_{OZ}$ and $L_{C}$.  

\begin{figure*}[ht]
\centering
  \includegraphics[height=2.2in]{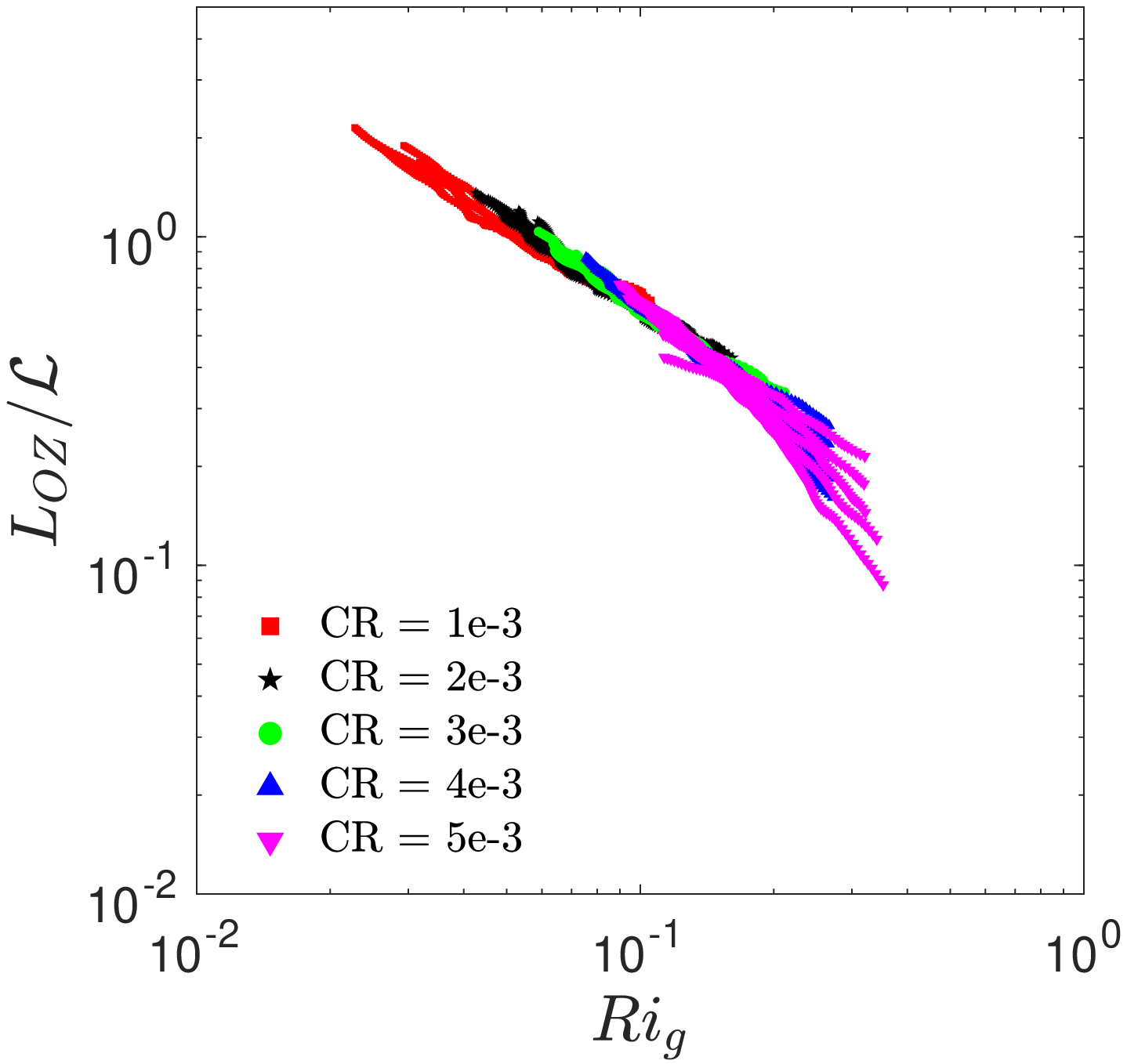}
  \hspace{0.3in}
  \includegraphics[height=2.2in]{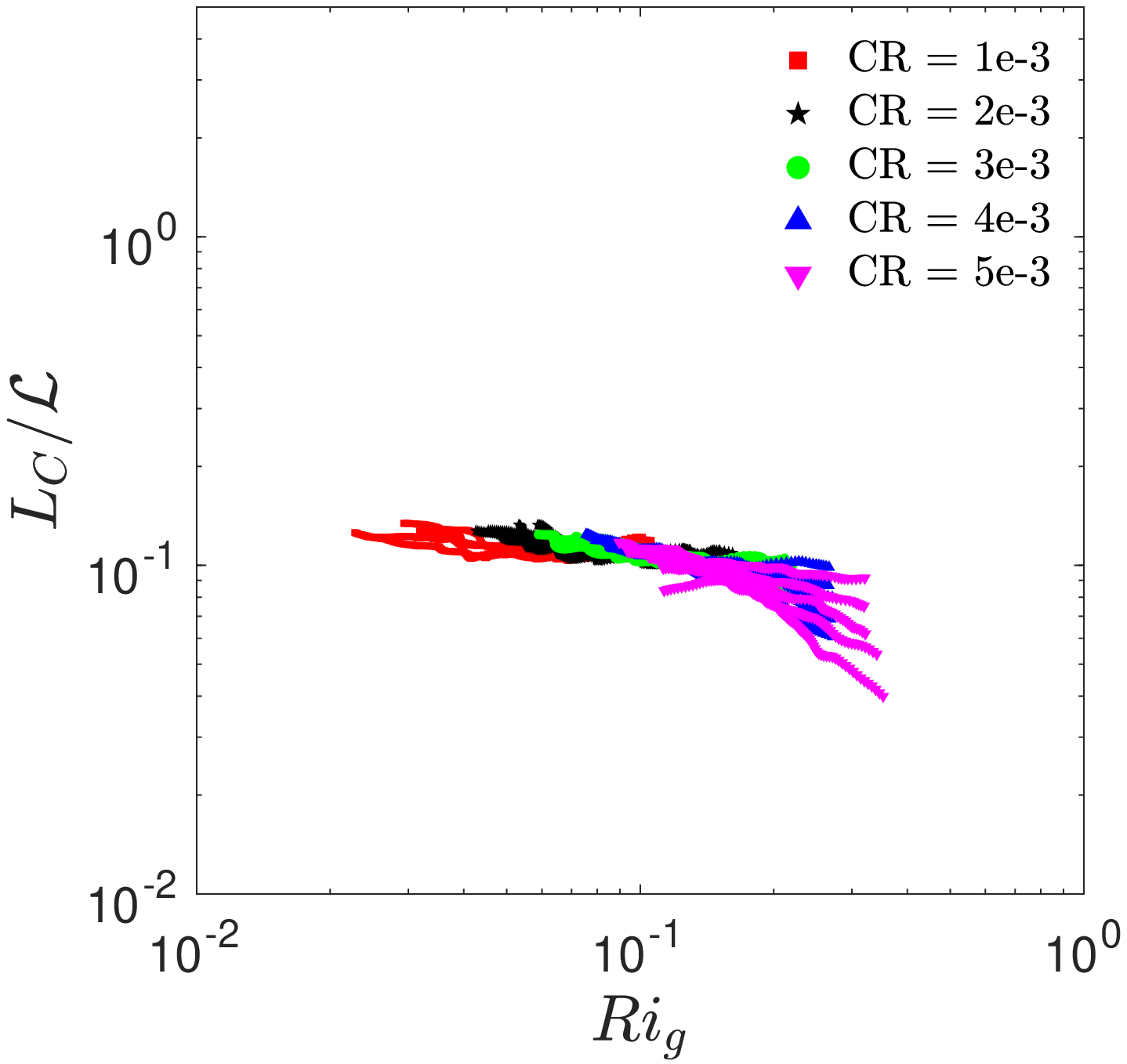}
\caption{Ozmidov (left panel) and Corrsin (right panel) length scales as functions of gradient Richardson numbers. These length scales are normalized by the integral length scale ($\mathcal{L}$). Simulated data from five different DNS runs are represented by different colored symbols in these plots. In the legends, $CR$ represents normalized cooling rates.}
\label{fig2}      
\end{figure*}

Since $\mathcal{L}$ changes across the simulations, the OLS values are normalized by corresponding $\mathcal{L}$ values and plotted as functions of $Ri_g$ in Fig.~\ref{fig2}. The collapse of the data from different runs, on to seemingly universal curves, is remarkable for all the cases except for $Ri_g > 0.2$. We would like to mention that similar scaling behavior was not found if other normalization factors (e.g., $h$) are used.       

Normalized $L_{OZ}$ decreases monotonically with $Ri_g$. In contrast, normalized $L_C$ barely exhibits any sensitivity to $Ri_g$ (except for $Ri_g > 0.1$). Even for weakly-stable condition, it is less than 20\% of $\mathcal{L}$. 
Based on the expressions of $L_{OZ}$, $L_C$ and $Ri_g$, we can write: 
\begin{equation}
    \frac{L_C}{L_{OZ}} = \left(\frac{N}{S}\right)^{3/2} = Ri_g^{3/4}.
\end{equation}
Thus, for $Ri_g < 1$, one expects $L_C < L_{OZ}$; this relationship is fully supported by Fig.~\ref{fig2}. In comparison to the buoyancy effects, the shear effects are felt at smaller length scales for the entire stability range considered in the present study. 

Dissipation rate of turbulent kinetic energy is used in the definitions for both $L_{OZ}$ and $L_{C}$. However, it is also possible to formulate OLSs based on the dissipation rate of temperature fluctuations as follows:
\begin{subequations}
\begin{equation}
    L_1 \equiv \left(\frac{g}{\Theta_0}\right)^{-1/4} \overline{\chi}^{1/2} \left(\frac{\partial\overline{\theta}}{\partial z} \right)^{-5/4},
    \label{EqL1}
\end{equation}
\begin{equation}
    L_2 \equiv \overline\varepsilon^{-1/4} \overline{\chi}^{3/4} \left(\frac{\partial\overline{\theta}}{\partial z} \right)^{-3/2},
    \label{EqL2}
\end{equation}
\begin{equation}
    L_3 \equiv \overline{\chi}^{1/2} \left(\frac{\partial\overline{\theta}}{\partial z} \right)^{-1} S^{-1/2},
    \label{EqL3}
\end{equation}
\begin{equation}
    L_4 \equiv \left(\frac{g}{\Theta_0}\right) \overline\chi^{1/2} S^{-5/2},
    \label{EqL4}
\end{equation}
\end{subequations}
where, $(\partial{\overline{\theta}}/\partial z)$ is the vertical gradient of mean potential temperature and $\Theta_0$ is a reference potential temperature. These length scales were proposed by Panchev based on dimensional analysis \citep{panchev75,monin85}. Characteristics of yet another OLS proposed by Bolgiano \cite{bolgiano59,bolgiano62} and Obukhov \cite{obukhov59} is discussed separately in Appendix 1. 

\begin{figure*}[ht]
\centering
  \includegraphics[height=2.2in]{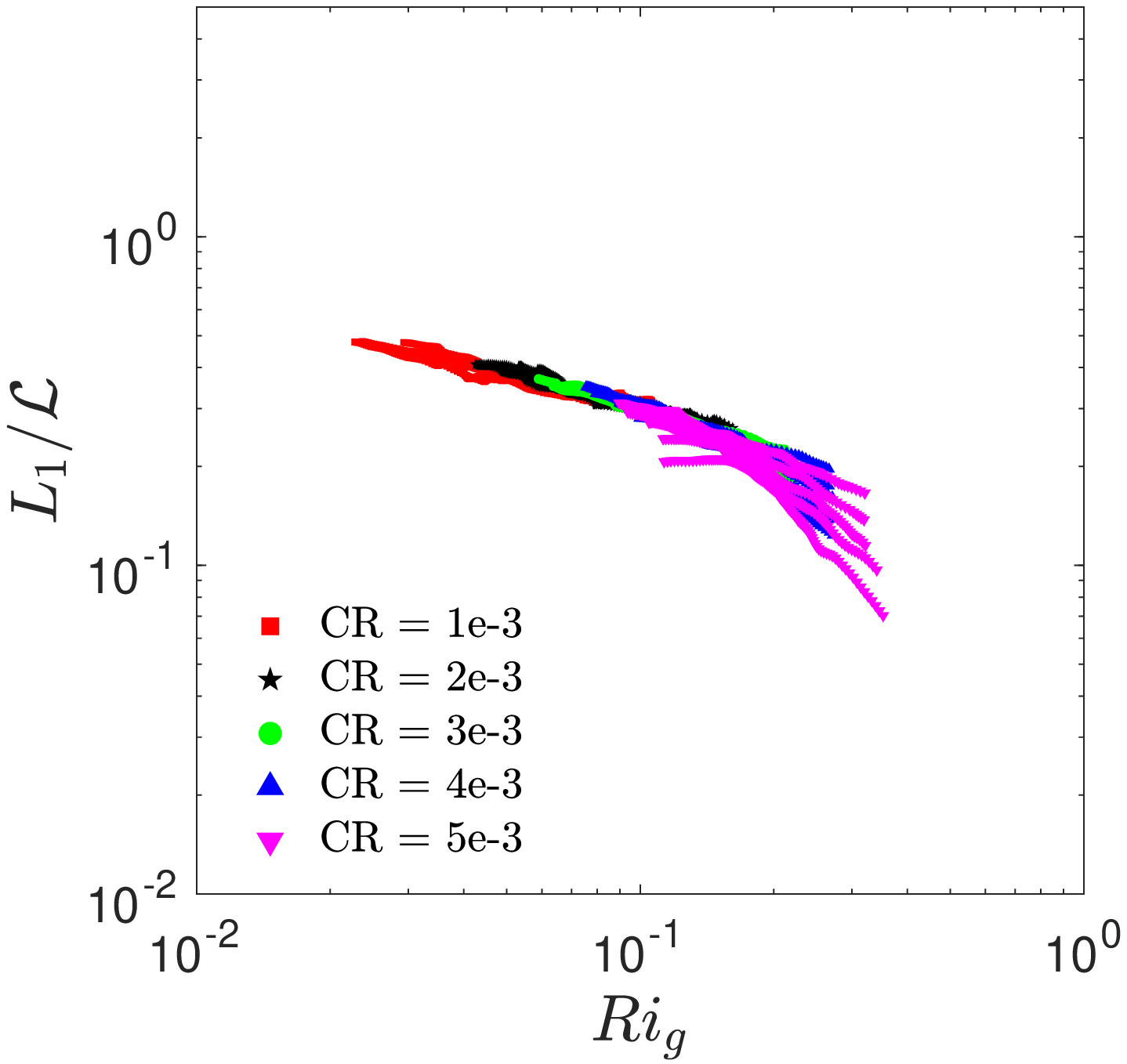}
  \hspace{0.3in}
  \includegraphics[height=2.2in]{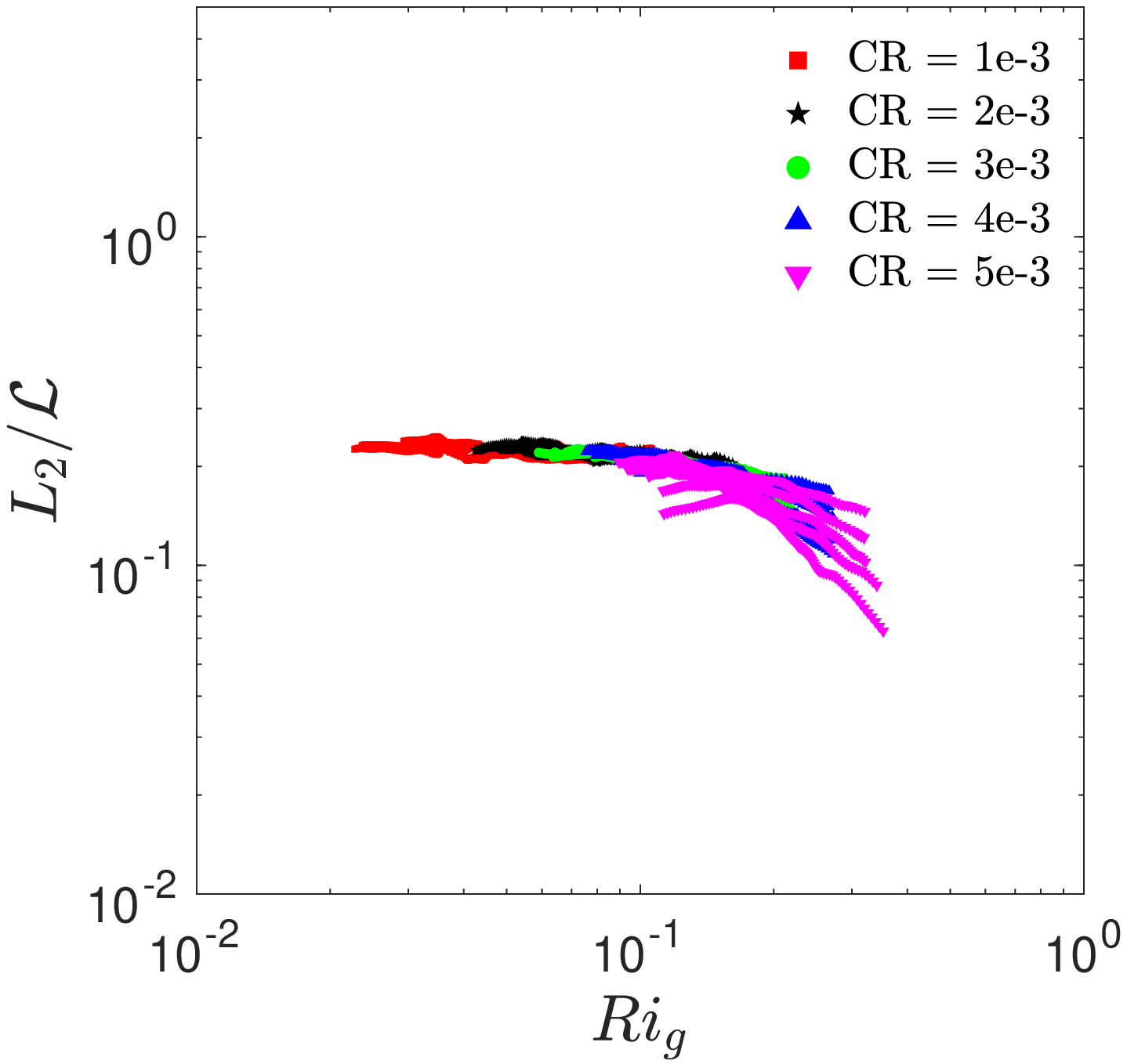}\\
  \includegraphics[height=2.2in]{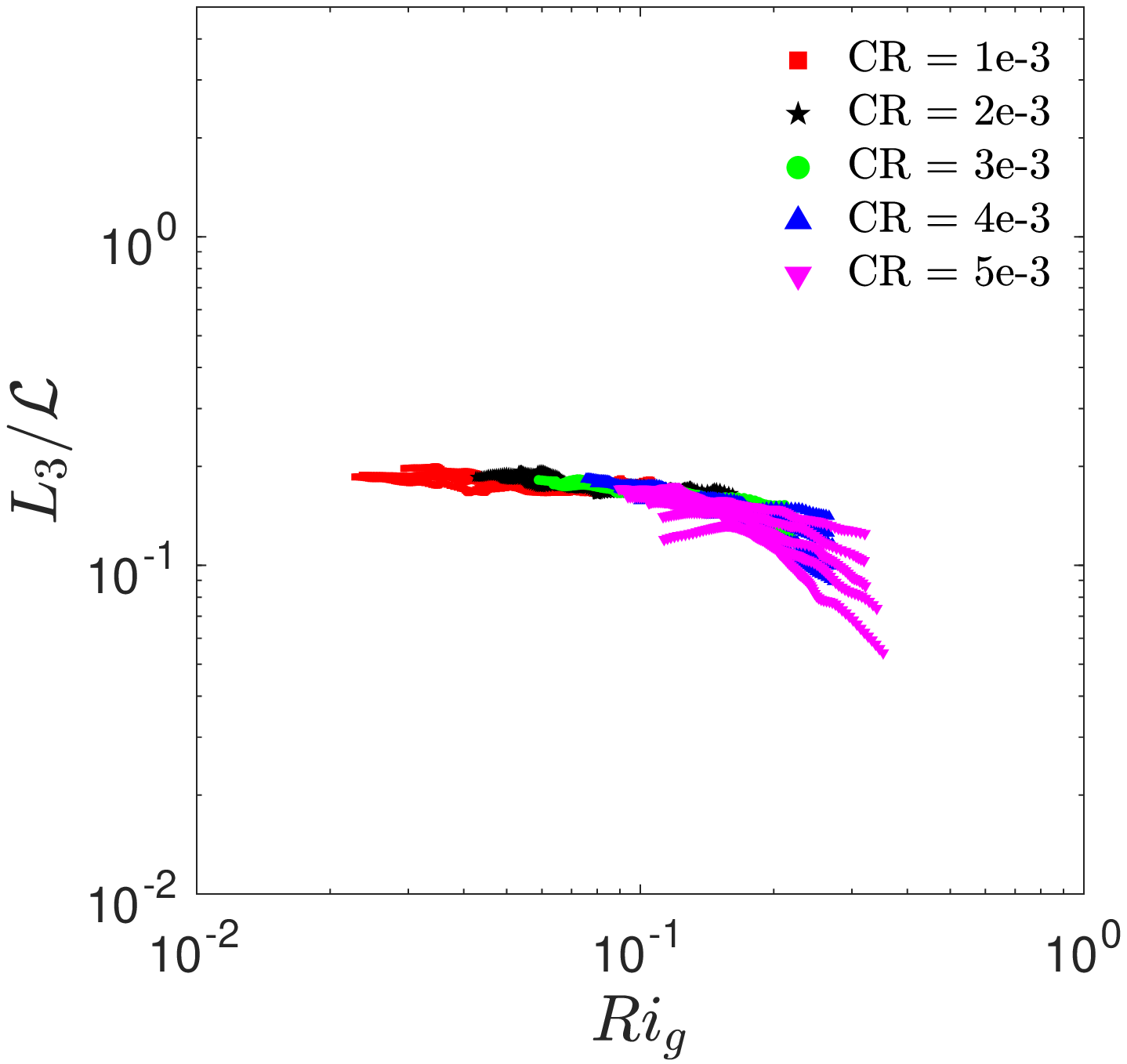}
  \hspace{0.3in}
  \includegraphics[height=2.2in]{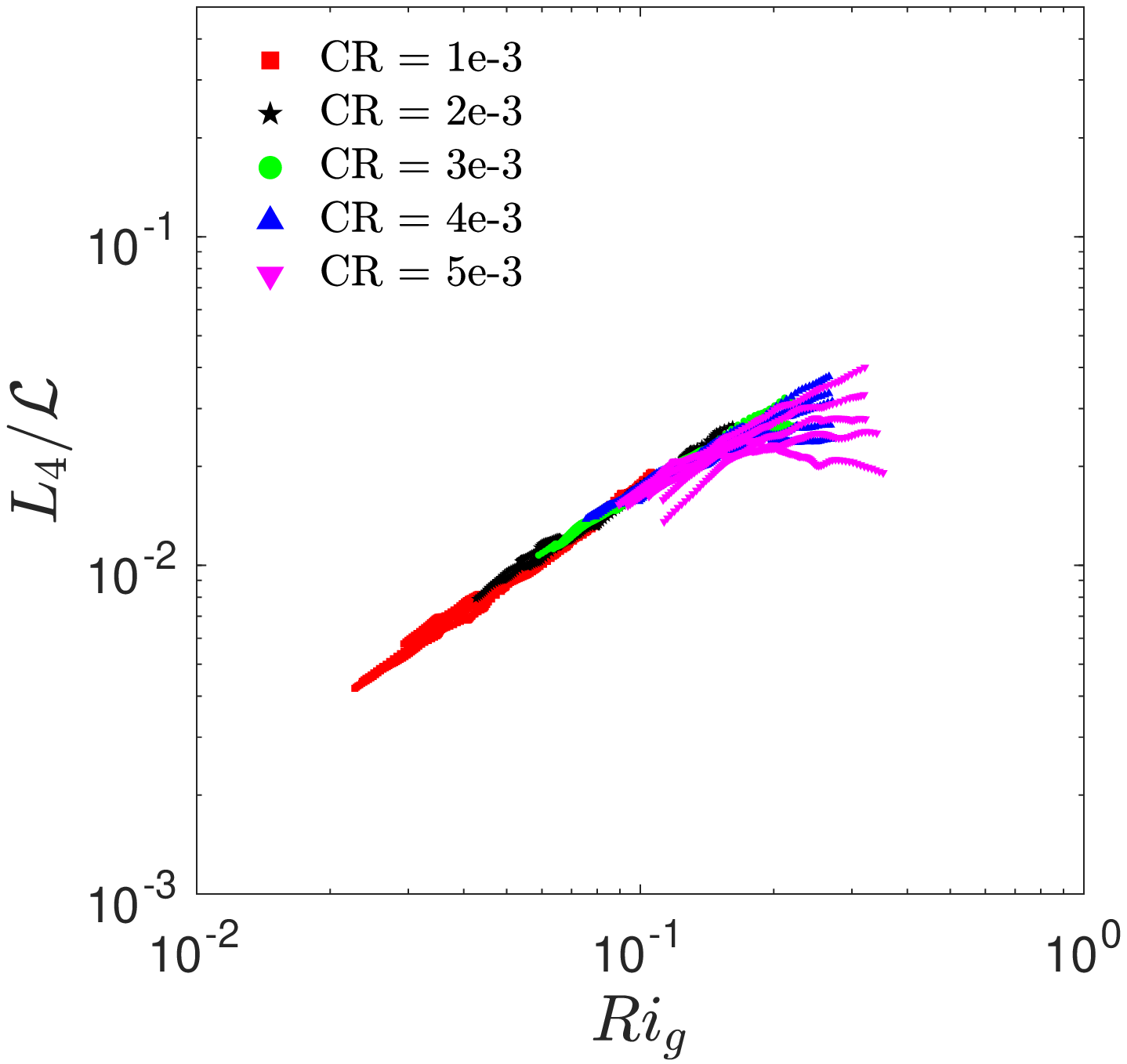}
\caption{$\chi$-based length scales as functions of gradient Richardson numbers. These length scales are normalized by the integral length scale. Simulated data from five different DNS runs are represented by different colored symbols in these plots. In the legends, $CR$ represents normalized cooling rates.}
\label{fig3}      
\end{figure*}

In Fig.~\ref{fig3}, the $\overline{\chi}$-based length scale formulations are plotted against $Ri_g$. Similar to $L_{OZ}$, the normalized $L_1$ monotonically decrease with increasing $Ri_g$. Whereas, the normalized $L_4$ increase with $Ri_g$ in an unphysical manner. It is quite evident that both the normalized $L_2$ and $L_3$ scales behave very similar to $L_C$ (see right panel of Fig.~\ref{fig2}).

\begin{figure*}[ht]
\centering
  \includegraphics[height=2.2in]{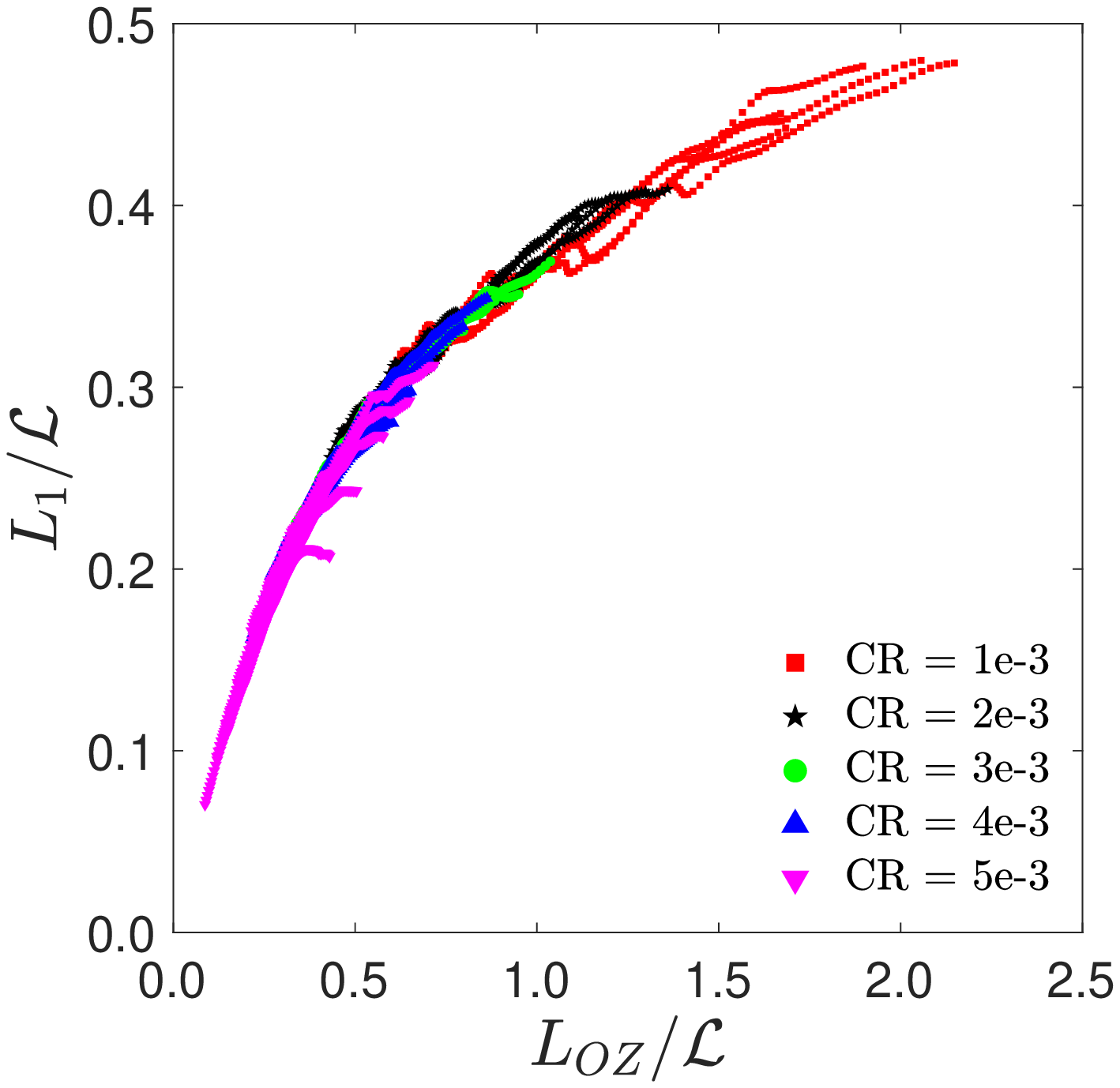} 
  \hspace{0.3in}
  \includegraphics[height=2.2in]{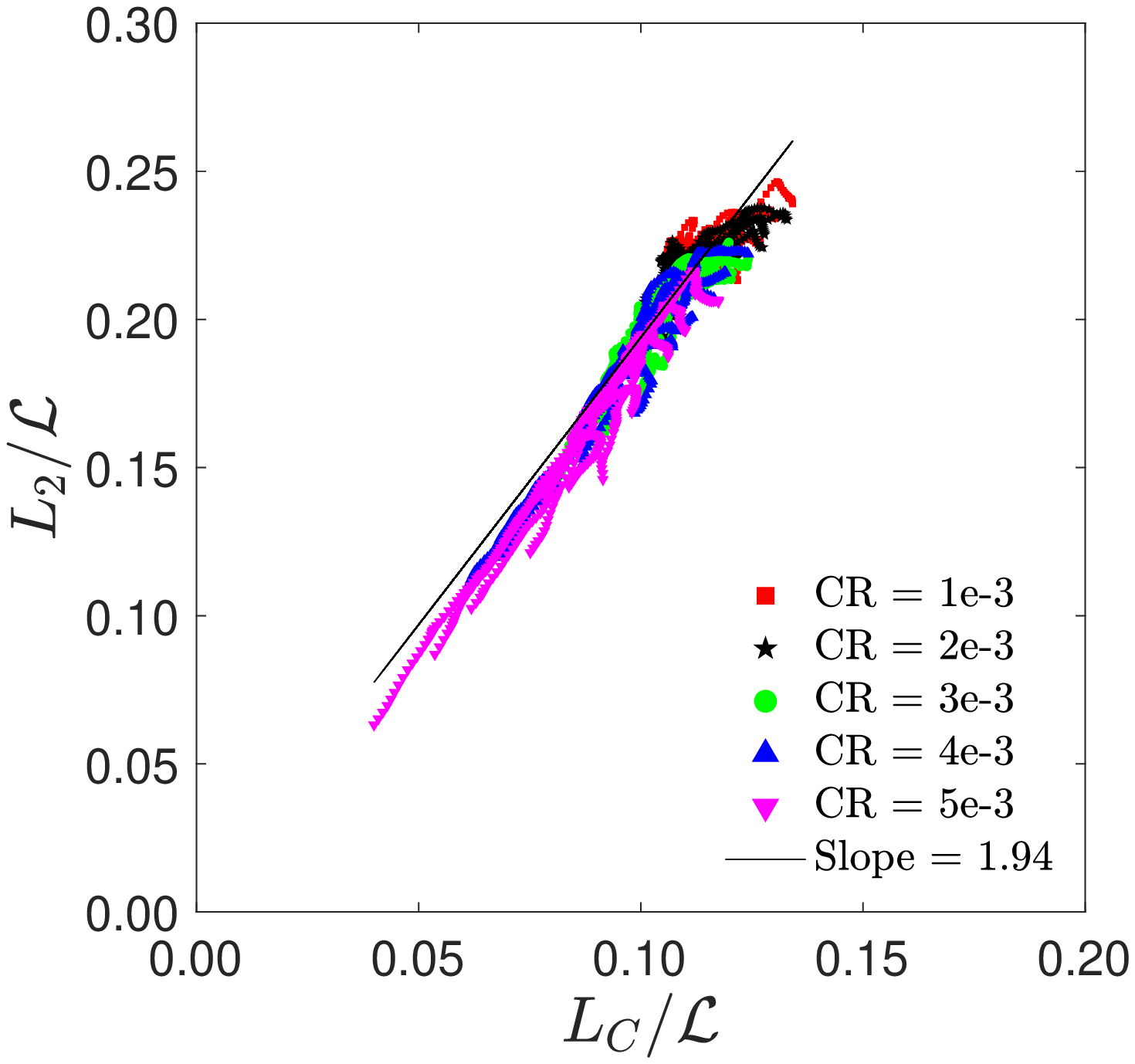}\\
  \includegraphics[height=2.2in]{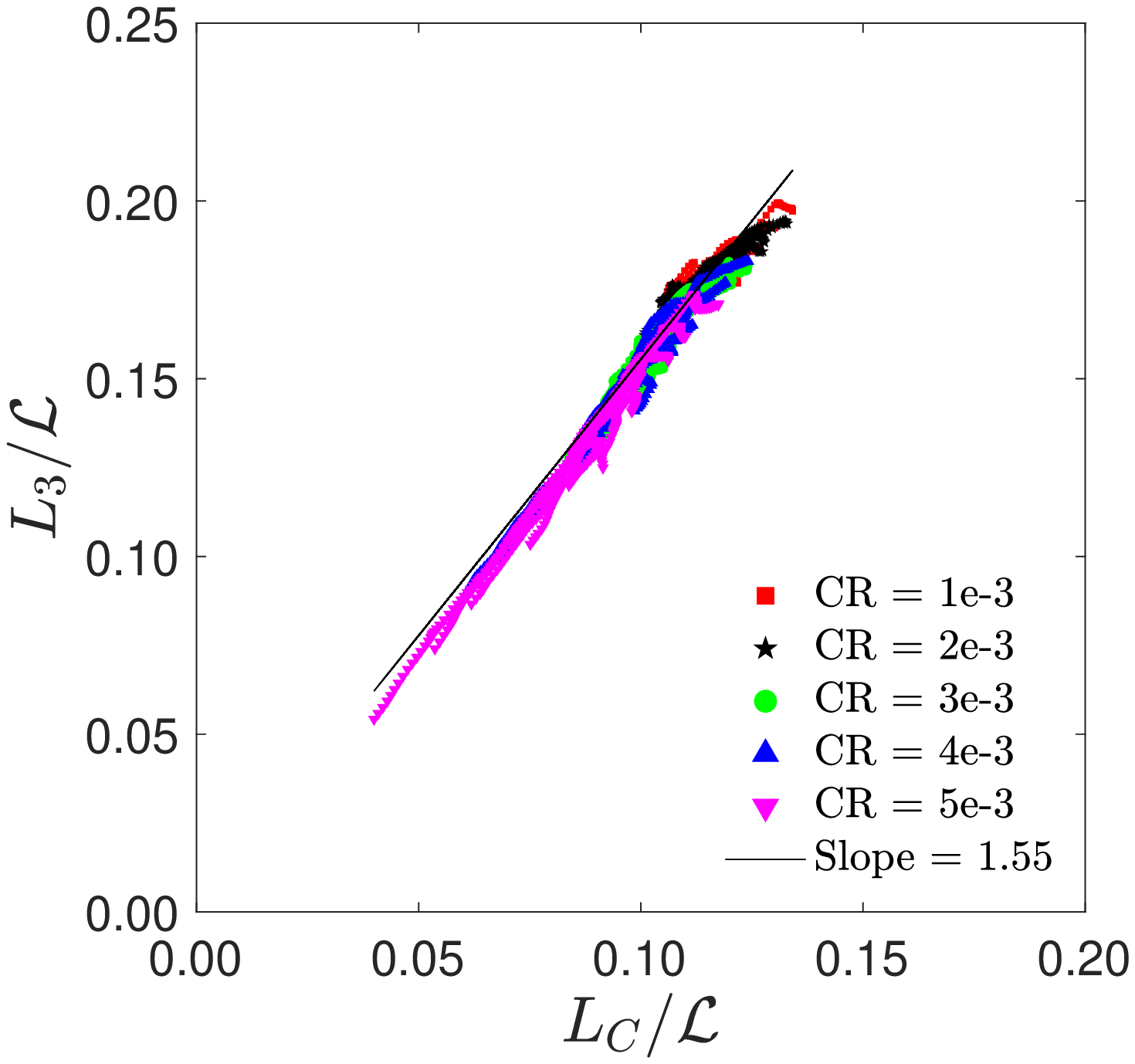}
  \hspace{0.3in}
  \includegraphics[height=2.2in]{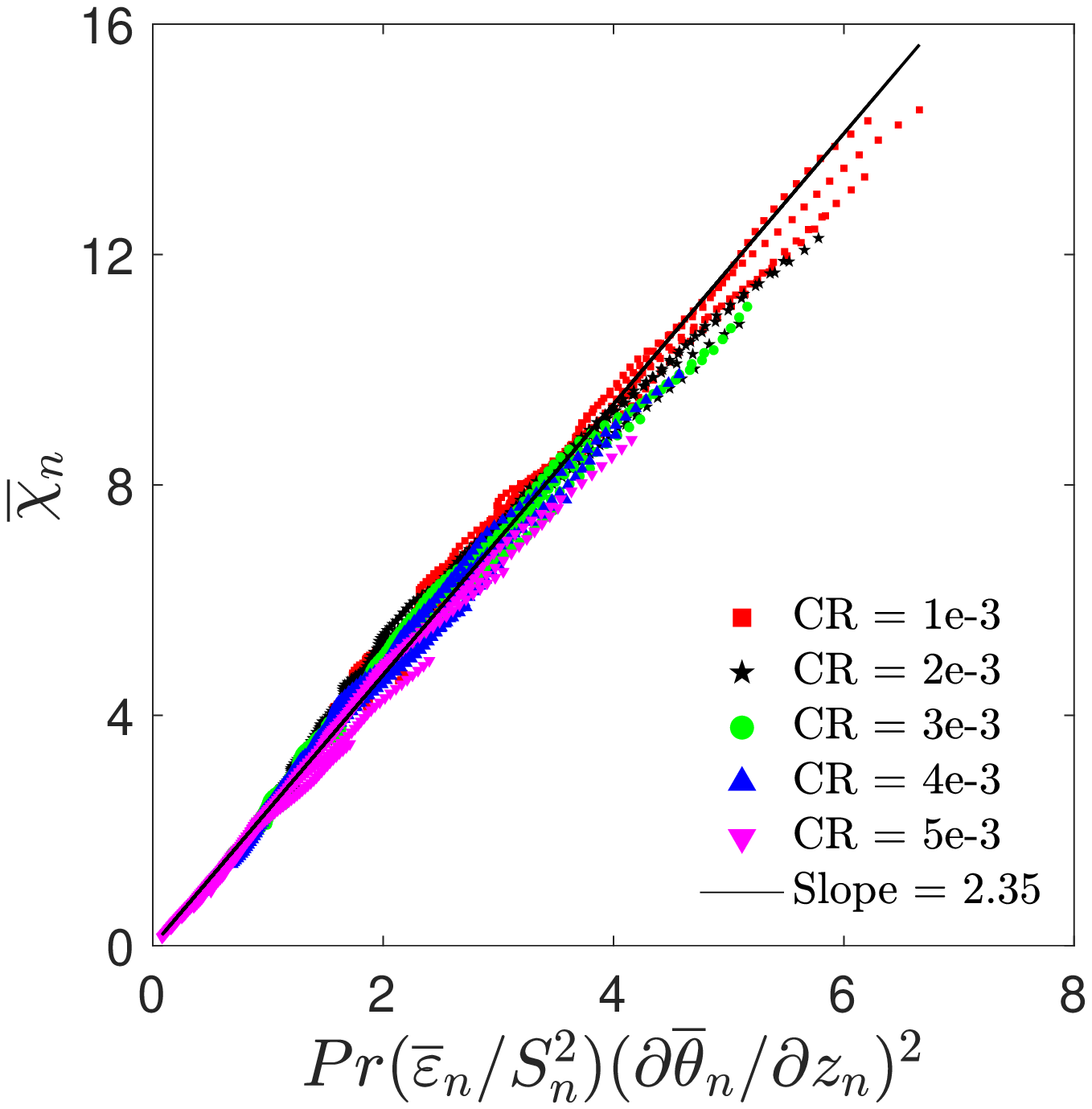}
  \caption{Variation of the normalized $\overline{\chi}$-based length scales against the normalized Ozmidov length scale (top-left panel) and the normalized Corrsin length scale (top-right and bottom-left panels). Bottom-right panel: normalized $\overline{\chi}$ as a function of normalized $\overline{\varepsilon}$, $S$, and $(\partial\overline{\theta}/\partial z)$. Please refer to Eq.~\ref{EDRCHI2}. Simulated data from five different DNS runs are represented by different colored symbols in these plots. In the legends, $CR$ represents normalized cooling rates.}
\label{fig4}      
\end{figure*}

Given the trends in Fig.~\ref{fig3}, we plotted a few inter-relationships of OLSs in Fig.~\ref{fig4}. In each case, the collapse of DNS-based data on a single curve is excellent. In the case of $L_1$-vs-$L_{OZ}$ plot, the curve is nonlinear. However, in the case of $L_2$-vs-$L_{C}$ and $L_3$-vs-$L_{C}$ plots, the data fall on more-or-less straight lines. The regressed slopes are reported in the legends of these plots. 

If we assume $L_2 \equiv L_C$, based on Eq.~\ref{LC} and Eq.~\ref{EqL2}, it is trivial to arrive at: 
\begin{equation}
    \frac{\overline{\chi}}{\overline{\varepsilon}} \approx \frac{\left(\frac{\partial\overline{\theta}}{\partial z} \right)^2}{S^2}.
    \label{EDRCHI2}
\end{equation}
Interestingly, the assumption of $L_3 \equiv L_C$ also leads to the same equation. As a matter of fact, this equation can be derived from the budget equations of TKE and temperature variance  with certain assumptions as elaborated below. Assuming steady-state condition, horizontal homogeneity, and neglecting the secondary terms (e.g., transport), we can write: 
\begin{subequations}
\begin{equation}
    \overline{\varepsilon} = -\overline{u'w'}\left(\frac{\partial \overline{u}}{\partial z}\right) -\overline{v'w'}\left(\frac{\partial \overline{v}}{\partial z}\right) + \left(\frac{g}{\Theta_0}\right)\overline{w'\theta'},
\end{equation}
and
\begin{equation}
    \overline{\chi} = -2 \overline{w'\theta'}\left(\frac{\partial \overline{\theta}}{\partial z}\right).
\end{equation}
\end{subequations}
If we apply K-theory, these equations can be further simplified to: 
\begin{subequations}
\begin{equation}
    \overline{\varepsilon} = K_M S^2 - K_H \left(\frac{g}{\Theta_0}\right) \left(\frac{\partial \overline{\theta}}{\partial z}\right),
    \label{EDRK}
\end{equation}
and
\begin{equation}
    \overline{\chi} = 2 K_H \left(\frac{\partial \overline{\theta}}{\partial z}\right)^2,
    \label{CHIK}
\end{equation}
\end{subequations}
where, $K_M$ and $K_H$ are eddy viscosity and diffusivity, respectively. By utilizing the definitions of gradient Richardson number ($Ri_g$) and turbulent Prandtl number ($Pr_T = K_M/K_H$), we can deduce from Eq.~\ref{EDRK} and Eq.~\ref{CHIK}: 
\begin{equation}
    \frac{\overline{\chi}}{\overline{\varepsilon}} =\frac{2}{\left(Pr_T - Ri_g\right)} \frac{\left(\frac{\partial\overline{\theta}}{\partial z} \right)^2}{S^2}.  
\end{equation}

Anderson~\cite{anderson09} conducted a rigorous statistical analysis of the observational data collected at the British Antarctic Survey's Halley station on the the Antarctic. He avoided the self-correlation issue and proposed the following empirical relationship for $0.01 < Ri_g < 0.25$: 
\begin{equation}
    Pr_T^{-1} = (0.84 \pm 0.03) Ri_g^{-0.105 \pm 0.012}
    \label{And09}
\end{equation}
Clearly, the $Ri_g$-dependence of the Prandtl number is rather weak for small values of $Ri_g$. Similar findings were reported in other experimental and modeling studies \citep[e.g.,][]{sukoriansky06,kantha18,li19}.

In the bottom-right panel of Fig.~\ref{fig4}, we have plotted Eq.~\ref{EDRCHI2} in a normalized form. The slope of the fitted line is 2.35. For $Ri_g = 0.2$, according to Eq.~\ref{And09}, $Pr_T \approx 1$. Thus, the ratio $2/(Pr_T - Ri_g)$ is approximately 2.48. When $Ri_g$ equals to 0.1, $Pr_T \approx 0.93$ following Eq.~\ref{And09}. In this case, the ratio $2/(Pr_T - Ri_g)$ is close to 2.40. These values are not far from the estimated slope of 2.35 in Fig.~\ref{fig4} (bottom-right panel). In other words, our DNS-based results are in-line with past observational studies.

\section{Structure Parameter of Temperature ($C_T^2$)}

Using the DNS database of the current study, Basu et al.~\cite{basu20} recently found that  $\overline{\varepsilon} = 0.23 \overline{e} S$ and $\overline{\varepsilon} = 0.63 \sigma_w^2 S$ for $0 < Ri_g < 0.2$. If we insert these formulations in Eq.~\ref{EDRCHI2}, we get:
\begin{subequations}
\begin{equation}
    \overline{\chi} \approx \left(\frac{\overline{e}}{S}\right) \left(\frac{\partial\overline{\theta}}{\partial z} \right)^2,
    \label{EDRCHI3}
\end{equation}
and
\begin{equation}
    \overline{\chi} \approx \left(\frac{\sigma_w^2}{S}\right) \left(\frac{\partial\overline{\theta}}{\partial z} \right)^2.
    \label{EDRCHI4}
\end{equation}
\label{EDRCHI34}
\end{subequations}

\begin{figure*}[ht]
\centering
  \includegraphics[height=2.2in]{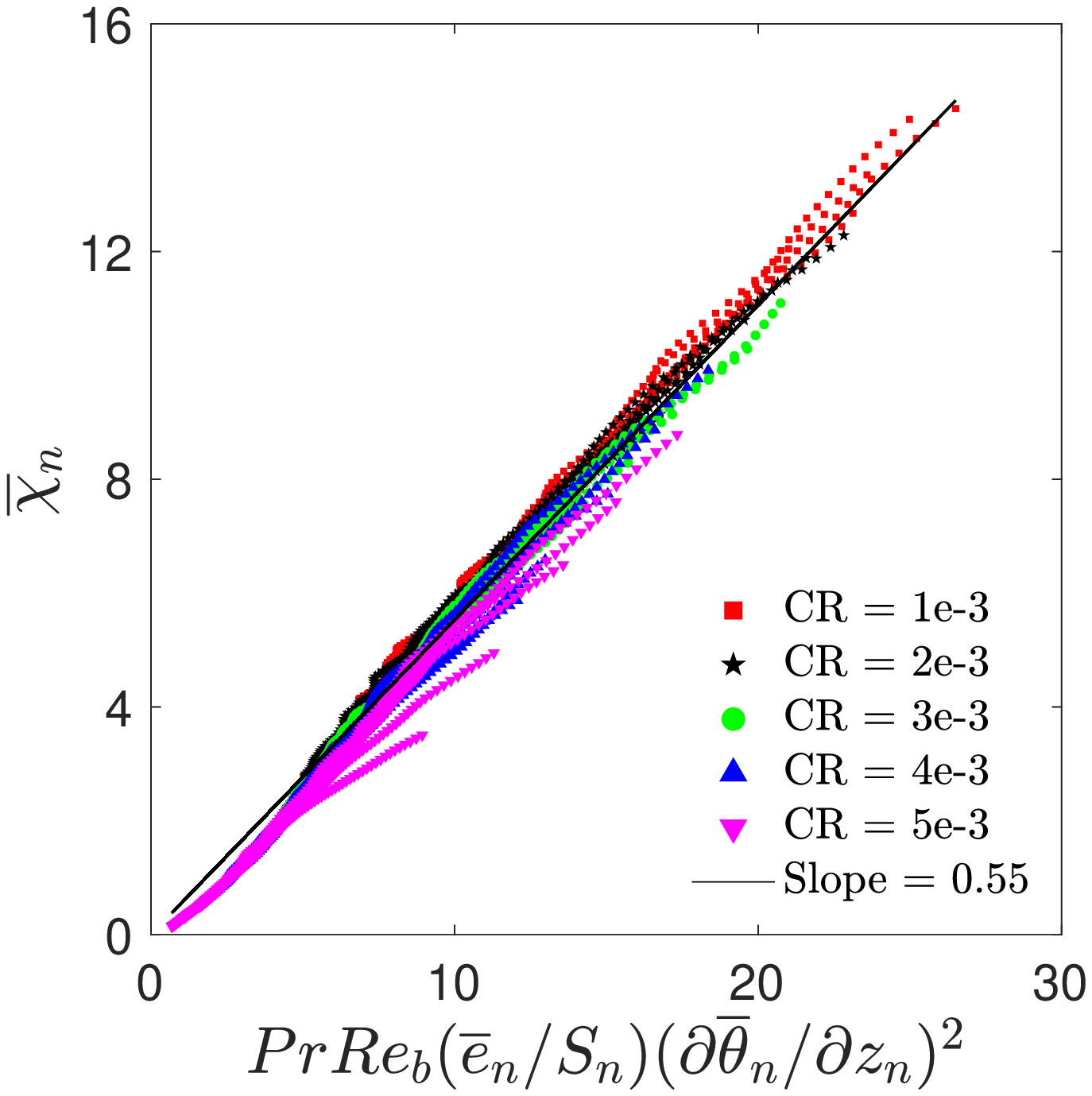}
  \hspace{0.3in}
  \includegraphics[height=2.2in]{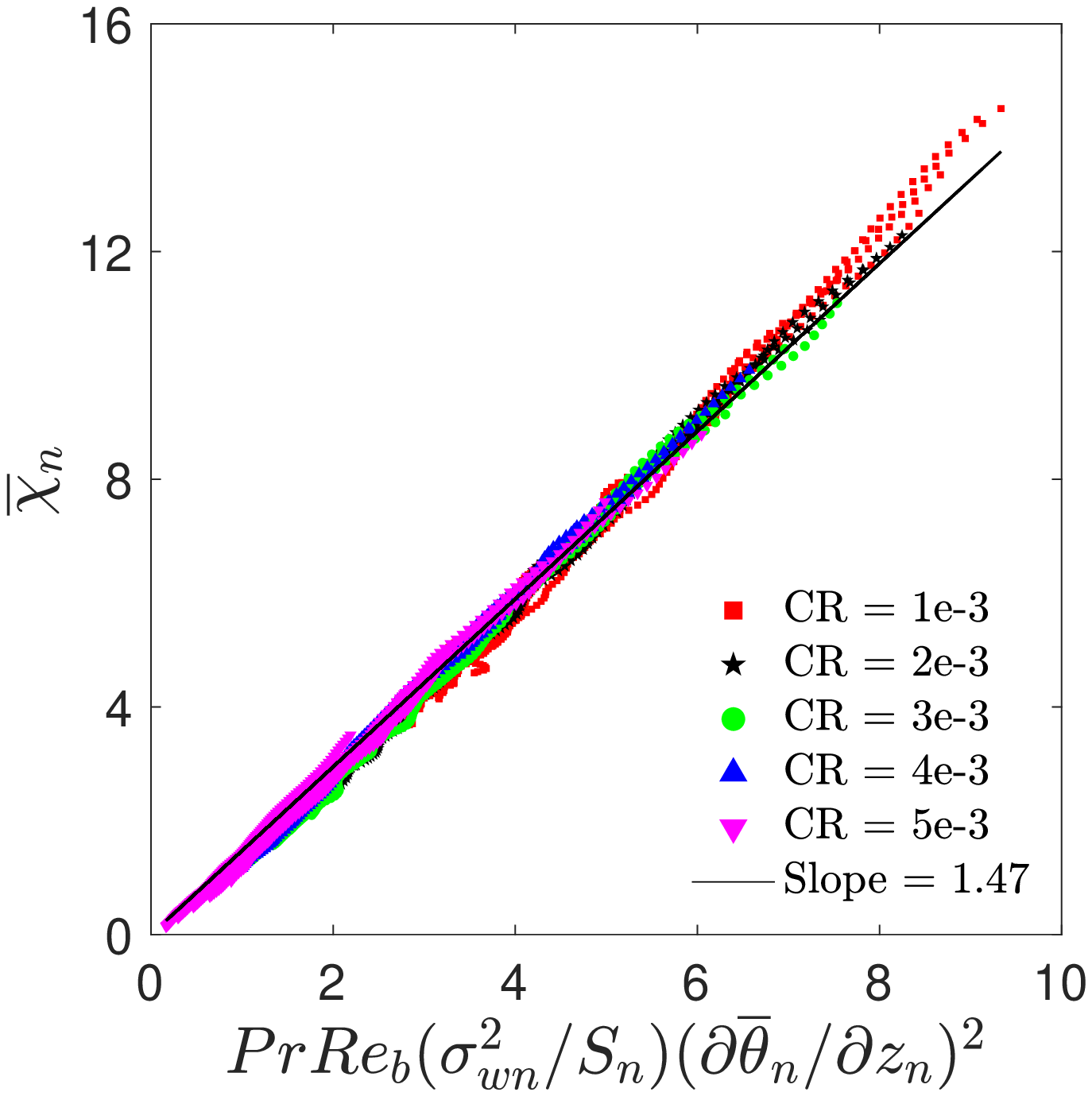}\\
  \includegraphics[height=2.2in]{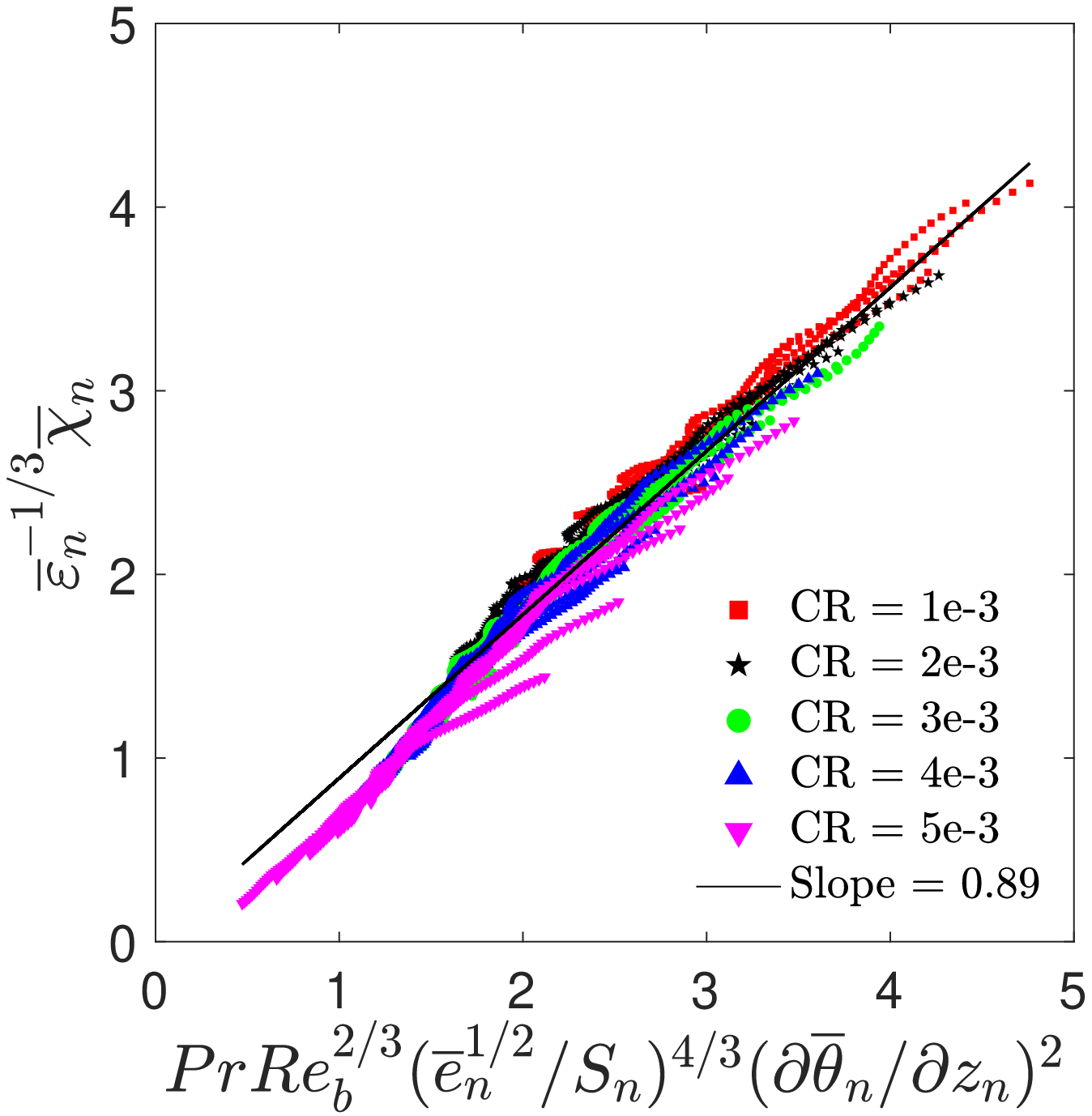}
  \hspace{0.3in}
  \includegraphics[height=2.2in]{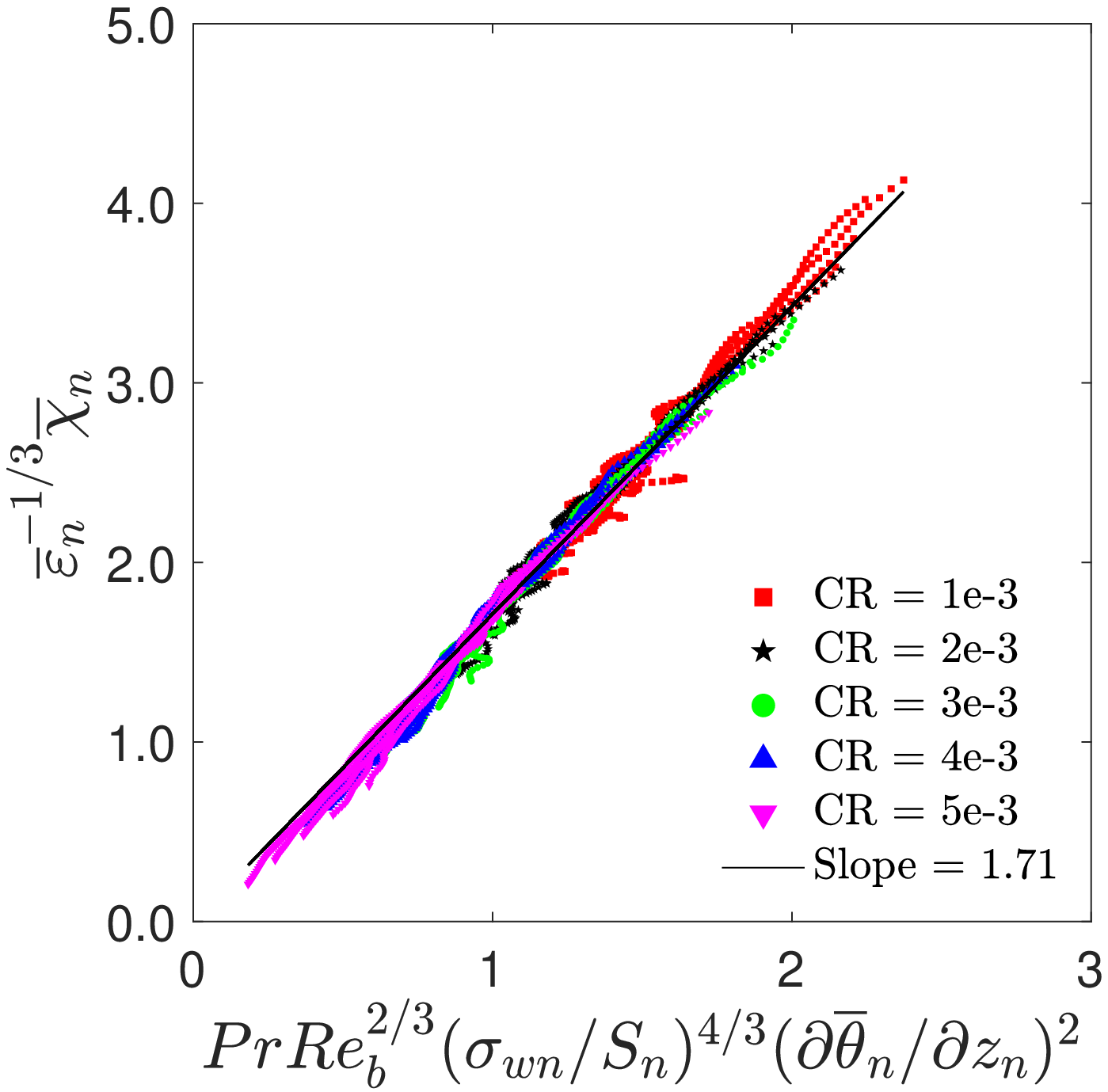}
\caption{Top panels: normalized $\overline{\chi}$ as a function of normalized $\overline{e}$, $\sigma_w^2$, $S$, and $(\partial\overline{\theta}/\partial z)$. Please refer to Eqs.~\ref{EDRCHI34}. Bottom panels: normalized $\overline{\varepsilon}^{-1/3}\overline{\chi}$ as a function of normalized $\overline{e}$, $\sigma_w^2$, $S$, and $(\partial\overline{\theta}/\partial z)$. Please refer to Eqs.~\ref{EqCT2}. Simulated data from five different DNS runs are represented by different colored symbols in these plots. In the legends, $CR$ represents normalized cooling rates.}
\label{fig5}      
\end{figure*}

\noindent The top panels of Fig.~\ref{fig5} strongly support the validity of these formulations. The proportionality constants in these equations are found to be equal to 0.55 and 1.47, respectively.

By definition, $C_T^2 \approx \overline{\varepsilon}^{-1/3} \overline{\chi}$. The proportionality constant is usually taken equal to 1.6 \citep{wyngaard71b,he16c}. Thus, we can write: 
\begin{subequations}
\begin{equation}
    C_T^2 \approx \overline{\varepsilon}^{-1/3} \overline{\chi} \approx \left(\frac{\overline{e}^{1/2}}{S}\right)^{4/3} \left(\frac{\partial\overline{\theta}}{\partial z} \right)^2,
    \label{EqCT2a}
\end{equation}
and
\begin{equation}
    C_T^2 \approx \overline{\varepsilon}^{-1/3} \overline{\chi} \approx \left(\frac{\sigma_w}{S}\right)^{4/3} \left(\frac{\partial\overline{\theta}}{\partial z} \right)^2
    \label{EqCT2b}
\end{equation}
\label{EqCT2}
\end{subequations}

\noindent In the bottom panels of Fig.~\ref{fig5}, we establish that these equations (especially Eq.~\ref{EqCT2b}) nicely hold for our DNS-generated data. 

Based on theoretical and numerical work, Hunt et al.~\cite{hunt88,hunt89} proposed the shear-based length scales, $L_H \equiv \left(\frac{\overline{e}^{1/2}}{S}\right)$ and $L_H \equiv \left(\frac{\sigma_w}{S}\right)$, as the characteristic length scales for $0 < Ri_g < 0.5$. Thus, we can re-write Eqs.~\ref{EqCT2} as: 
\begin{equation}
    C_T^2 \approx L_H^{4/3} \left(\frac{\partial\overline{\theta}}{\partial z} \right)^2.
    \label{CT2_LH}
\end{equation}
A very similar equation was proposed by Tatarskii more than 50 years ago \citep{tatarski61,tatarskii71}, albeit with an OLS which needs to be prescribed. In the literature, several empirical parameterizations were proposed for this unknown length scale \citep{vanzandt78,coulman88,dewan93,basu15}. In this study, based on DNS-generated data, we demonstrate that the outer length scale in Tatarskii's equation should be equal to $L_H$ for $0 < Ri_g < 0.2$. 

At this point, we point out an interesting relationship that one can further deduce from our findings. If we compare Eq.~\ref{EDRCHI} against Eq.~\ref{EDRCHI2}, we get: 
\begin{equation}
\frac{\sigma_\theta^2}{\overline{e}} \approx \frac{\left(\frac{\partial\overline{\theta}}{\partial z} \right)^2}{S^2}.
\label{F76}
\end{equation}

\begin{figure*}[ht]
\centering
  \includegraphics[height=2.2in]{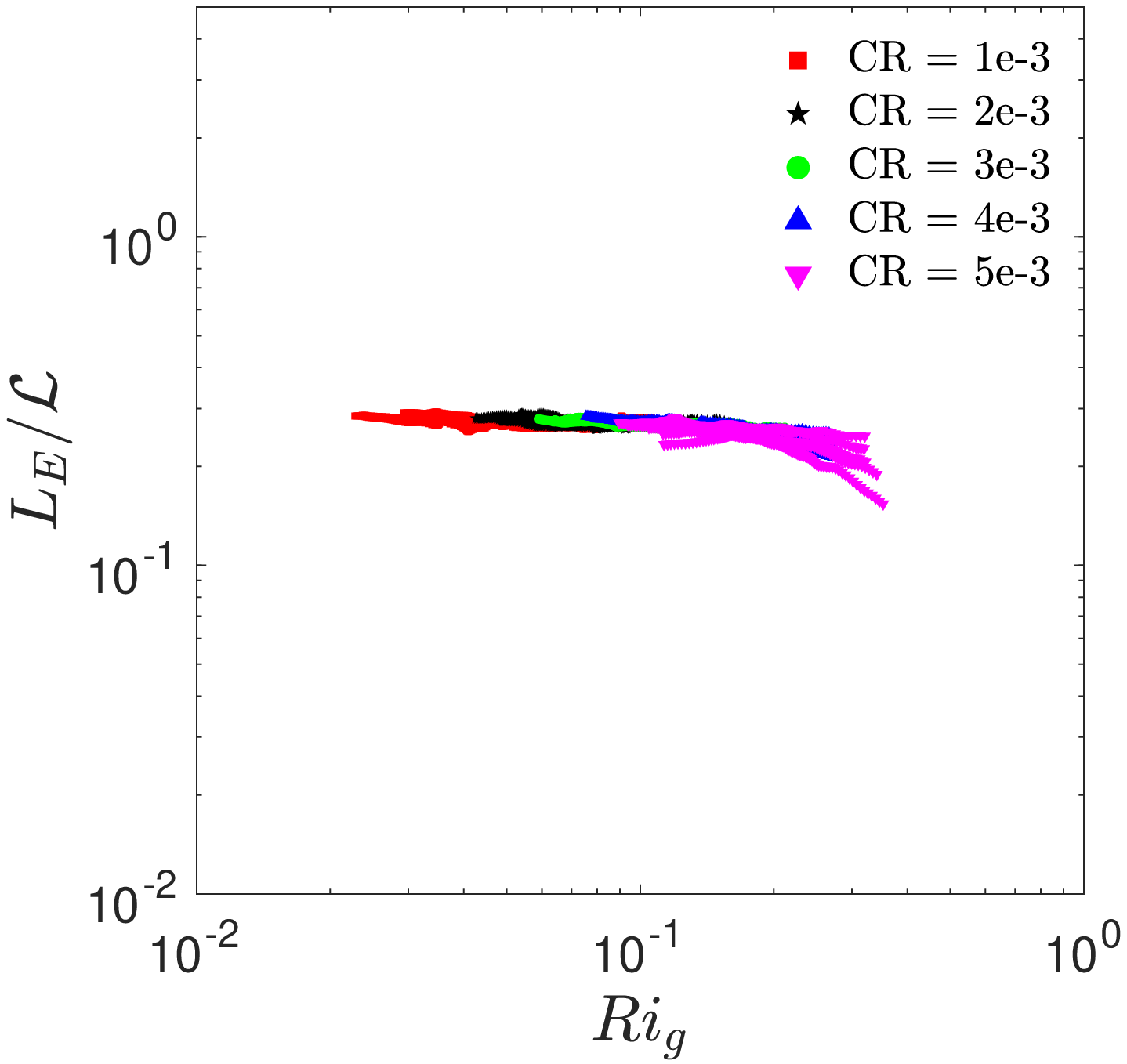}
  \hspace{0.3in}
  \includegraphics[height=2.2in]{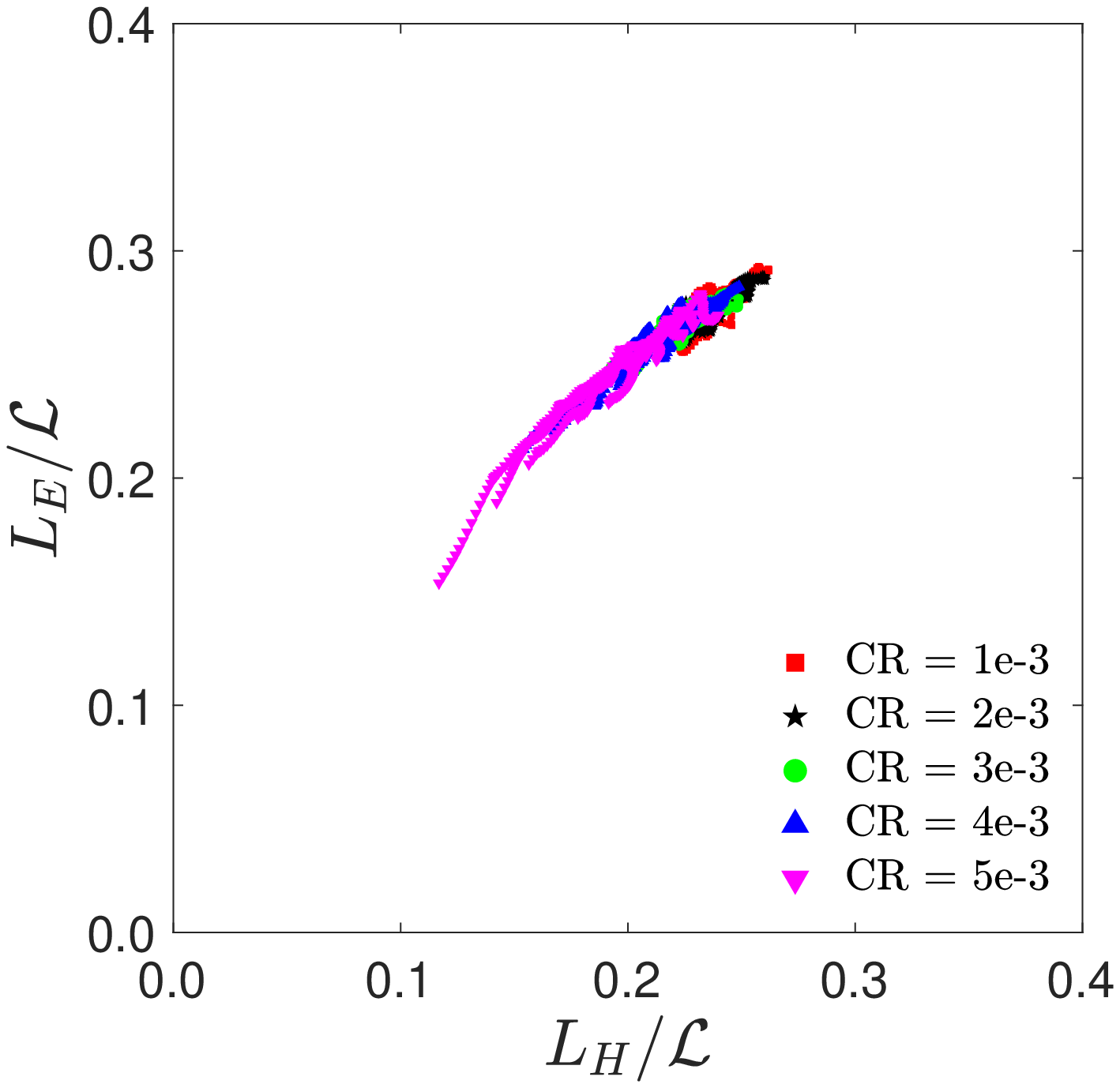}
\caption{Left panel: normalized Ellison length scale as a function of gradient Richardson number. Right panel: normalized Ellison length scale as a function of normalized Hunt length scale. Simulated data from five different DNS runs are represented by different colored symbols in these plots. In the legends, $CR$ represents normalized cooling rates.}
\label{fig6}      
\end{figure*}

Equivalently, one can write: 
\begin{subequations}
\begin{equation}
\frac{\sigma_\theta}{\left(\frac{\partial\overline{\theta}}{\partial z} \right)} \approx \frac{\overline{e}^{1/2}}{S},     
\end{equation}
or,
\begin{equation}
L_E \equiv L_H.     
\end{equation}
\end{subequations}
where, $L_E$ is a length scale proposed by Ellison~\cite{ellison57}. The dependence of $L_E$ on $Ri_g$ is documented in the left panel of Fig.~\ref{fig6}. In the right panel, we show the one-to-one relationship between $L_E$ and $L_H$. With the exception of a few data points from the simulation with the strongest cooling rate, it is clear that these length scales are linearly related to each other. Thus, the following equation can be used as a viable alternative to Eq.~\ref{CT2_LH}: 
\begin{equation}
    C_T^2 \approx L_E^{4/3} \left(\frac{\partial\overline{\theta}}{\partial z} \right)^2.
    \label{CT2_LE}
\end{equation}

In the literature, several studies have demonstrated the similarities between the so-called Thorpe scale \citep[$L_T$;][]{thorpe77,thorpe05} and $L_E$ using observed and simulated data \cite[e.g.,][]{itsweire93,mater13}. A simple heuristic derivation was also provided by \cite{gavrilov05}. Thus, it is plausible to replace $L_E$ with $L_T$ in Eq.~\ref{CT2_LE}: 
\begin{equation}
    C_T^2 \approx L_T^{4/3} \left(\frac{\partial\overline{\theta}}{\partial z} \right)^2.
    \label{CT2_LT}
\end{equation}
This equation was proposed by \cite{basu15} and was validated using observational data from a field campaign over Mauna Kea, Hawaii. 

In summary, we conjecture that Eqs.~\ref{CT2_LH}, ~\ref{CT2_LE}, and ~\ref{CT2_LT} are all valid parameterizations for $C_T^2$ as long as $Ri_g$ does not exceed 0.2. For larger values of $Ri_g$, a different length scale might be more appropriate; our present DNS runs cannot shed light on such a strong stability regime. 

\section{Concluding Remarks}

In this study, we analyze DNS-generated data to characterize several integral and outer length scales. From these results, we propose simple parameterizations for $\chi$ and $C_T^2$ when gradient Richardson number is less than 0.2. In the continuously turbulent atmospheric stable boundary layer, $Ri_g$ is usually less than 0.2 \citep{garratt82,nieuwstadt84}. Thus, the proposed parameterizations should be suitable for certain practical boundary layer problems. However, they will have limited applications for intermittently stable conditions.

In closing, we would like to emphasize the importance of Eq.~\ref{F76}. To the best of our knowledge, it was first reported by Fulachier and Dumas~\cite{fulachier76} from boundary layer experiments over a slightly heated plate. In a latter study, Fulachier and Antonia~\cite{fulachier84} found this formulation to hold for various other types of flows. They even concluded: 
\begin{quote}
    ``It seems therefore reasonable, from both mathematical and physical points of view, to seek a relationship, not between momentum and heat fluxes, as in the case with the Reynolds analogy, but preferably between the turbulent kinetic energy and the temperature variance.''
\end{quote}
To the best of our knowledge, Eq.~\ref{F76} is not used in atmospheric boundary layer studies. Since our findings are in agreement, we strongly endorse the assertion of \cite{fulachier84} and advocate further research on this equation.  

\section*{Data and Code Availability}
The DNS code (HERCULES) is available from: \url{https://github.com/friedenhe/HERCULES}. All the analysis codes and processed data are publicly available at \url{https://doi.org/10.5281/zenodo.3992818}. Given the sheer size of the raw DNS dataset, it will not be uploaded on to any repository; however, it will be available upon request from the authors.  

\begin{acknowledgements}
The first author thanks Bert Holtslag for having interesting discussions on this topic. The authors acknowledge computational resources obtained from the Department of Defense Supercomputing Resource Center (DSRC) for the direct numerical simulations. The views expressed in this paper do not reflect official policy or position by the U.S Air Force or the U.S. Government.
\end{acknowledgements}

\section*{Appendix 1: Bolgiano--Obukhov Length Scale}

Bolgiano~\cite{bolgiano59,bolgiano62} and Obukhov~\cite{obukhov59} independently proposed a buoyancy-range scaling and the following OLS based on theoretical arguments: 
\begin{equation}
    L_{BO} \equiv \left(\frac{g}{\Theta_0} \right)^{-3/2} \overline{\varepsilon}^{5/4} \overline{\chi}^{-3/4}
\end{equation}
Several laboratory and numerical studies \cite[e.g.,][]{niemela00,boffetta12} reported the existence of Bolgiano-Obukhov scaling in unstable condition. However, studies involving stably stratified conditions are rather limited \citep{rosenberg15,alam19}. Recently, Kumar et al.~\cite{kumar14} and Verma~\cite{verma18} reported that the Bolgiano-Obukhov scaling only exists for moderately stable condition. It is non-existent for near-neutral and very stable conditions.

\begin{figure*}[ht]
\centering
  \includegraphics[height=2.2in]{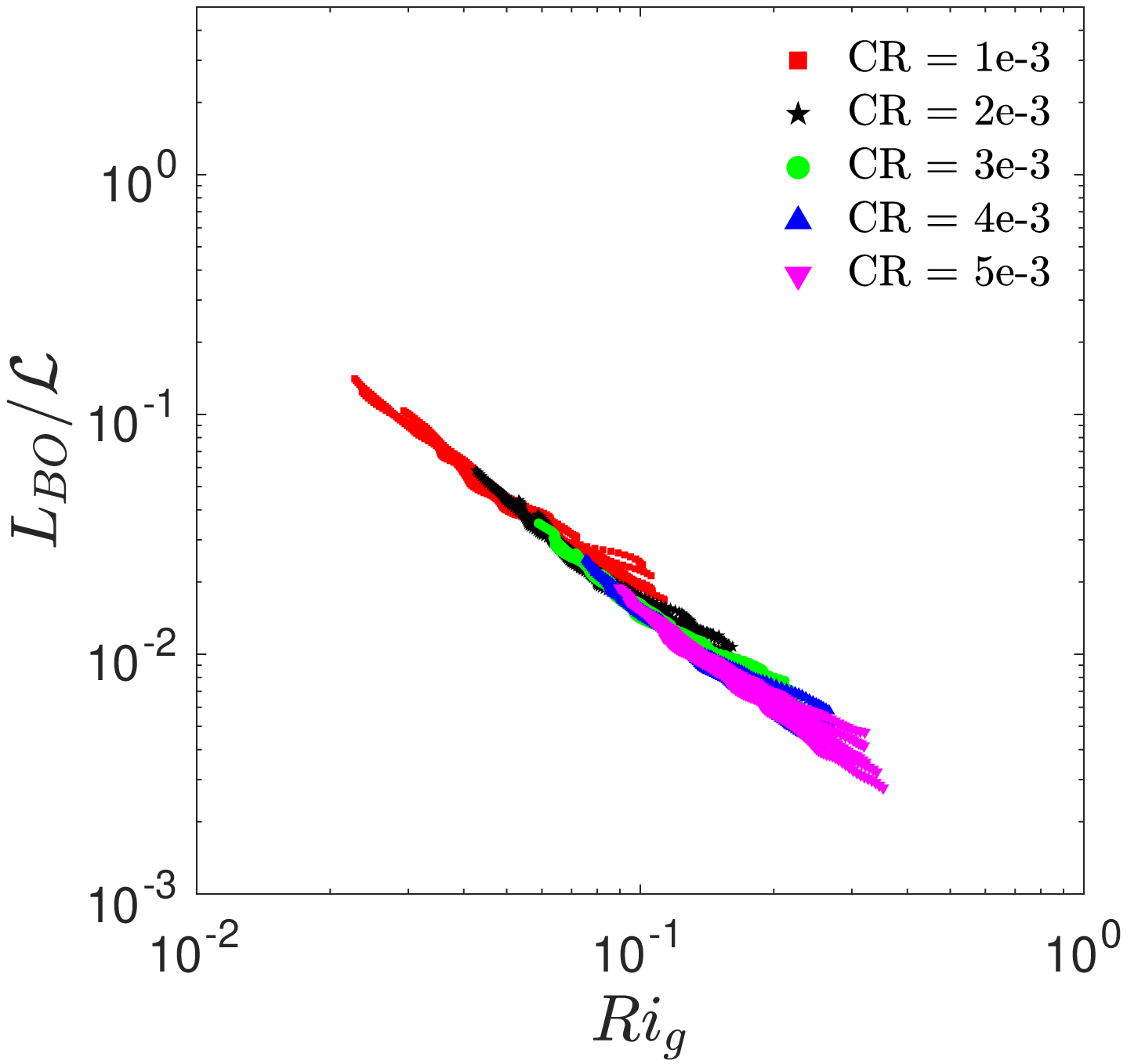}
  \hspace{0.3in}
  \includegraphics[height=2.2in]{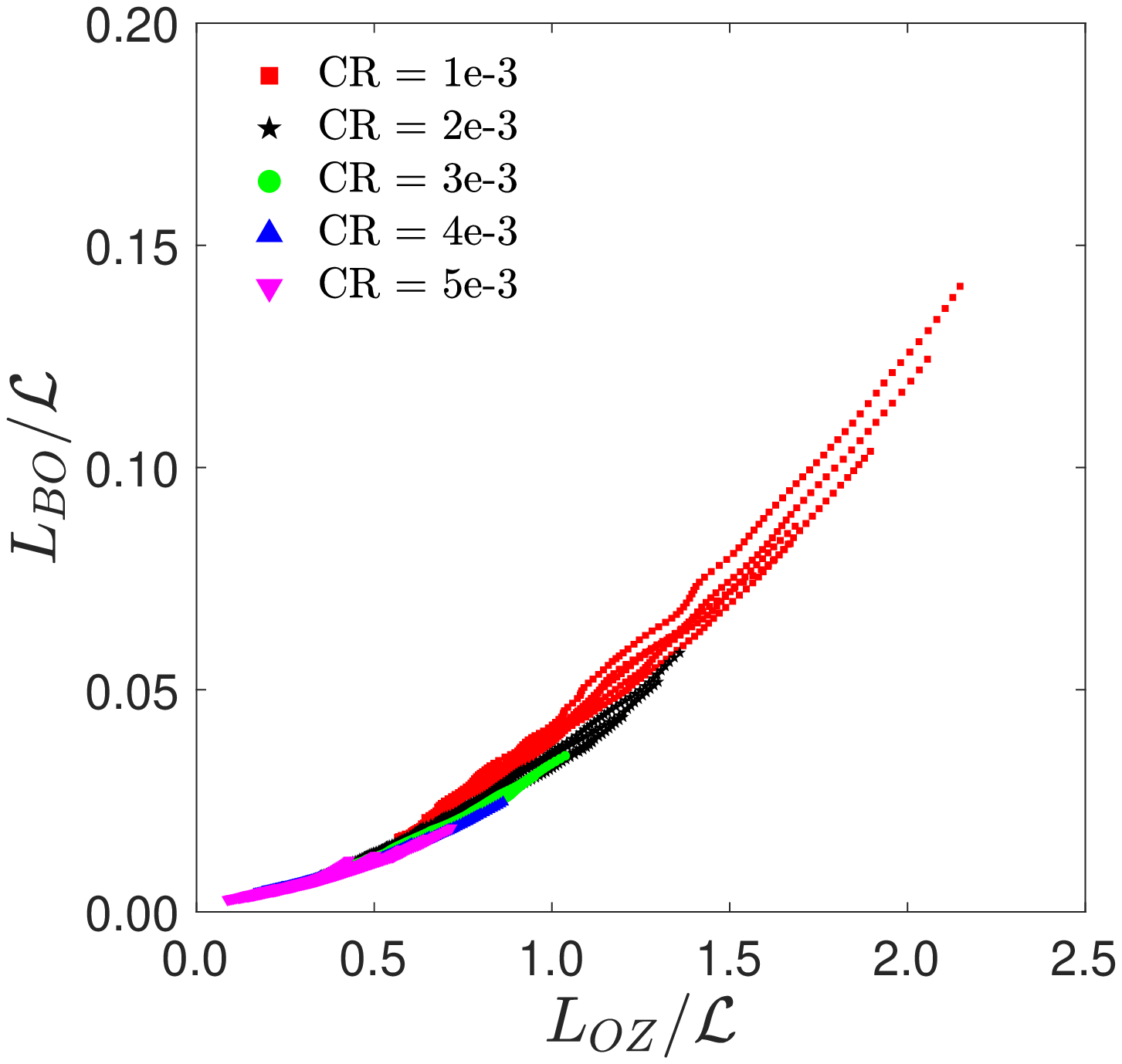}
\caption{Left panel: normalized Bolgiano length scale as a function of gradient Richardson number. Right panel: normalized Bolgiano length scale as a function of normalized Ozmidov length scale. Simulated data from five different DNS runs are represented by different colored symbols in these plots. In the legends, $CR$ represents normalized cooling rates.}
\label{fig7}      
\end{figure*}

In Fig.~\ref{fig7}, we show the traits of $L_{BO}$ as a function of $Ri_g$. Similar to $L_{OZ}$ and $L_1$, this length scale also shows a decreasing trend with increasing stability. However, the relationship between $L_{BO}$ and $L_{OZ}$ is nonlinear. As a consequence, we were unable to derive any simple formulation involving $\overline{\varepsilon}$, $\overline{\chi}$, and other variables. 

\section*{Appendix 2: Normalization of DNS Variables}

In DNS, the relevant variables are normalized as follows: 
\begin{subequations}
\begin{equation}
z_n = \frac{z}{h}, 
\end{equation}
\begin{equation}
u_n = \frac{u}{U_b}, 
\end{equation}
\begin{equation}
v_n = \frac{v}{U_b}, 
\end{equation}
\begin{equation}
w_n = \frac{w}{U_b}, 
\end{equation}
\begin{equation}
\theta_n = \frac{\theta-\Theta_{top}}{\Theta_{top}-\Theta_{bot}}. 
\end{equation}
\end{subequations}
After differentiation, we get: 
\begin{subequations}
\begin{equation}
\frac{\partial u}{\partial z} =  \frac{\partial u}{\partial z_n} \frac{\partial z_n}{\partial z} = \frac{\partial u}{\partial u_n} \frac{\partial u_n}{\partial z_n}\frac{\partial z_n}{\partial z} = \frac{U_b}{h} \frac{\partial u_n}{\partial z_n}, 
\end{equation}
\begin{equation}
\frac{\partial v}{\partial z} =  \frac{\partial v}{\partial z_n} \frac{\partial z_n}{\partial z} = \frac{\partial v}{\partial v_n} \frac{\partial v_n}{\partial z_n}\frac{\partial z_n}{\partial z} = \frac{U_b}{h} \frac{\partial v_n}{\partial z_n}, 
\end{equation}
\begin{equation}
S =  \sqrt{\left(\frac{\partial \overline{u}}{\partial z}\right)^2 + \left(\frac{\partial \overline{v}}{\partial z} \right)^2} = \frac{U_b}{h} S_n,
\label{EqS}
\end{equation}
\begin{widetext}
\begin{equation}
\frac{\partial \theta}{\partial z} =  \frac{\partial \theta}{\partial z_n} \frac{\partial z_n}{\partial z} = \frac{\partial \theta}{\partial \theta_n}
\frac{\partial \theta_n}{\partial z_n}\frac{\partial z_n}{\partial z} = \left(\frac{\Theta_{top}-\Theta_{bot}}{h}\right) \frac{\partial \theta_n}{\partial z_n}. 
\label{EqdTdz}
\end{equation}
\end{widetext}
\end{subequations}
\begin{widetext}
The gradient Richardson number can be expanded as: 
\begin{equation}
    Ri_g = \frac{N^2}{S^2} = \frac{\left(\frac{g}{\Theta_0}\right)\left(\frac{\partial \overline{\theta}}{\partial z}\right)}{S^2}
    = \left(\frac{g}{\Theta_{top}}\right) \left(\frac{\Theta_{top}-\Theta_{bot}}{h}\right)\left(\frac{h}{U_b} \right)^2 \frac{\left(\frac{\partial \overline{\theta}_n}{\partial z_n} \right)}{S_n^2}.
\end{equation}
\end{widetext}
Using the definition of $Ri_b$ (see Sect.~2), we re-write $Ri_g$ as follows:
\begin{equation}
    Ri_g = Ri_b \frac{\left(\frac{\partial \overline{\theta}_n}{\partial z_n} \right)}{S_n^2}.
\end{equation}
Similarly, $N^2$ can be written as: 
\begin{equation}
    N^2 = Ri_b \left(\frac{U_b^2}{h^2}\right) \left(\frac{\partial\overline{\theta}_n}{\partial z_n}\right).
    \label{EqN2}
\end{equation}
The velocity variances, TKE, and temperature variance can be normalized as: 
\begin{subequations}
\begin{equation}
\sigma_{u_n}^2 = \frac{\sigma_u^2}{U_b^2},
\end{equation}
\begin{equation}
\sigma_{v_n}^2 = \frac{\sigma_v^2}{U_b^2},
\end{equation}
\begin{equation}
\sigma_{w_n}^2 = \frac{\sigma_w^2}{U_b^2},
\label{EqSigw}
\end{equation}
\begin{equation}
\overline{e}_n = \frac{\overline{e}}{U_b^2}, 
\label{Eqe}
\end{equation}
\begin{equation}
\sigma_{\theta_n}^2 = \frac{\sigma_\theta^2}{\left(\Theta_{top}-\Theta_{bot}\right)^2}.
\label{EqSigT}
\end{equation}
\end{subequations}
Following the above normalization approach, we can also derive the following relationships for the dissipation rate of TKE and variance of temperature fluctuations:
\begin{subequations}
\begin{equation}
    \overline{\varepsilon} = \nu \left(\frac{U_b}{h}\right)^2 \overline{\varepsilon}_n. 
    \label{EqEDR}
\end{equation}
\begin{equation}
    \overline{\chi} = k \left(\frac{\Theta_{top}-\Theta_{bot}}{h} \right)^2 \overline{\chi}_n
    \label{EqCHI}
\end{equation}
\end{subequations}
We can combine Eqs.~\ref{Eqe}, \ref{EqSigT}, \ref{EqEDR}, and \ref{EqCHI}, we can re-write Eq.~\ref{EDRCHI} as follows:
\begin{equation}
    \overline{\chi}_n = \left(\frac{\nu}{k}\right) \left(\frac{\sigma_{\theta_n}^2}{\overline{e}_n}\right) \overline{\varepsilon}_n = Pr \left(\frac{\sigma_{\theta_n}^2}{\overline{e}_n}\right) \overline{\varepsilon}_n.
\end{equation}
In a similar fashion, we can utilize Eqs.~\ref{EqS}, \ref{EqdTdz}, \ref{Eqe}, and \ref{EqCHI} to re-write Eq.~\ref{EDRCHI3} as follows:
\begin{widetext}
\begin{equation}
    \overline{\chi}_n = \left(\frac{\nu}{k}\right) \left(\frac{U_b h}{\nu} \right) \left(\frac{\overline{e}_n}{S_n} \right) \left( \frac{\partial\overline{\theta}_n}{\partial z_n}\right)^2 = Pr Re_b \left(\frac{\overline{e}_n}{S_n} \right) \left( \frac{\partial\overline{\theta}_n}{\partial z_n}\right)^2.
\end{equation}
\end{widetext}

\section*{Appendix 3: Supplementary Analyses of DNS-generated Data}

\begin{figure*}[ht]
\centering
  \includegraphics[height=2.2in]{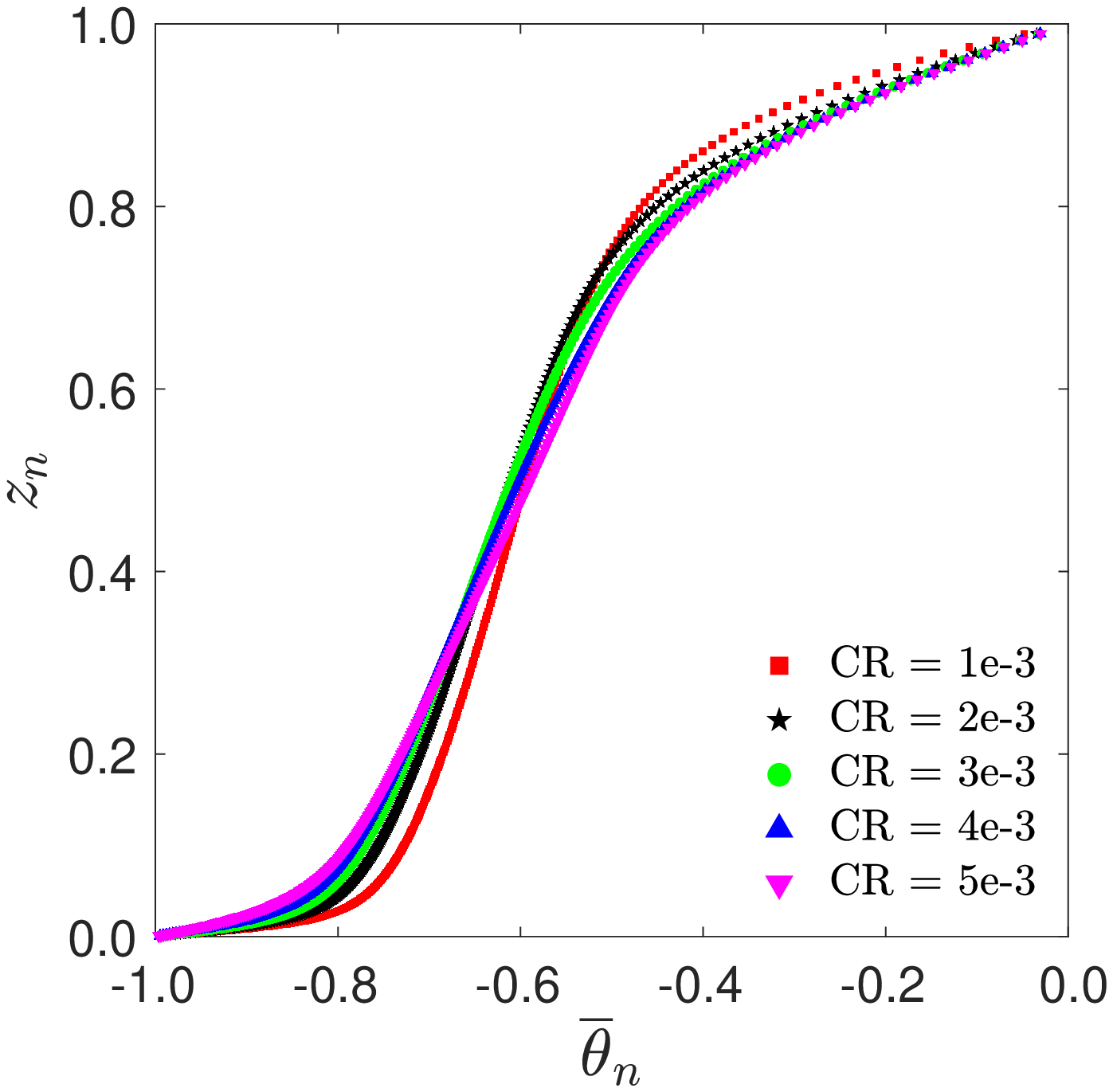}
  \hspace{0.3in}
  \includegraphics[height=2.2in]{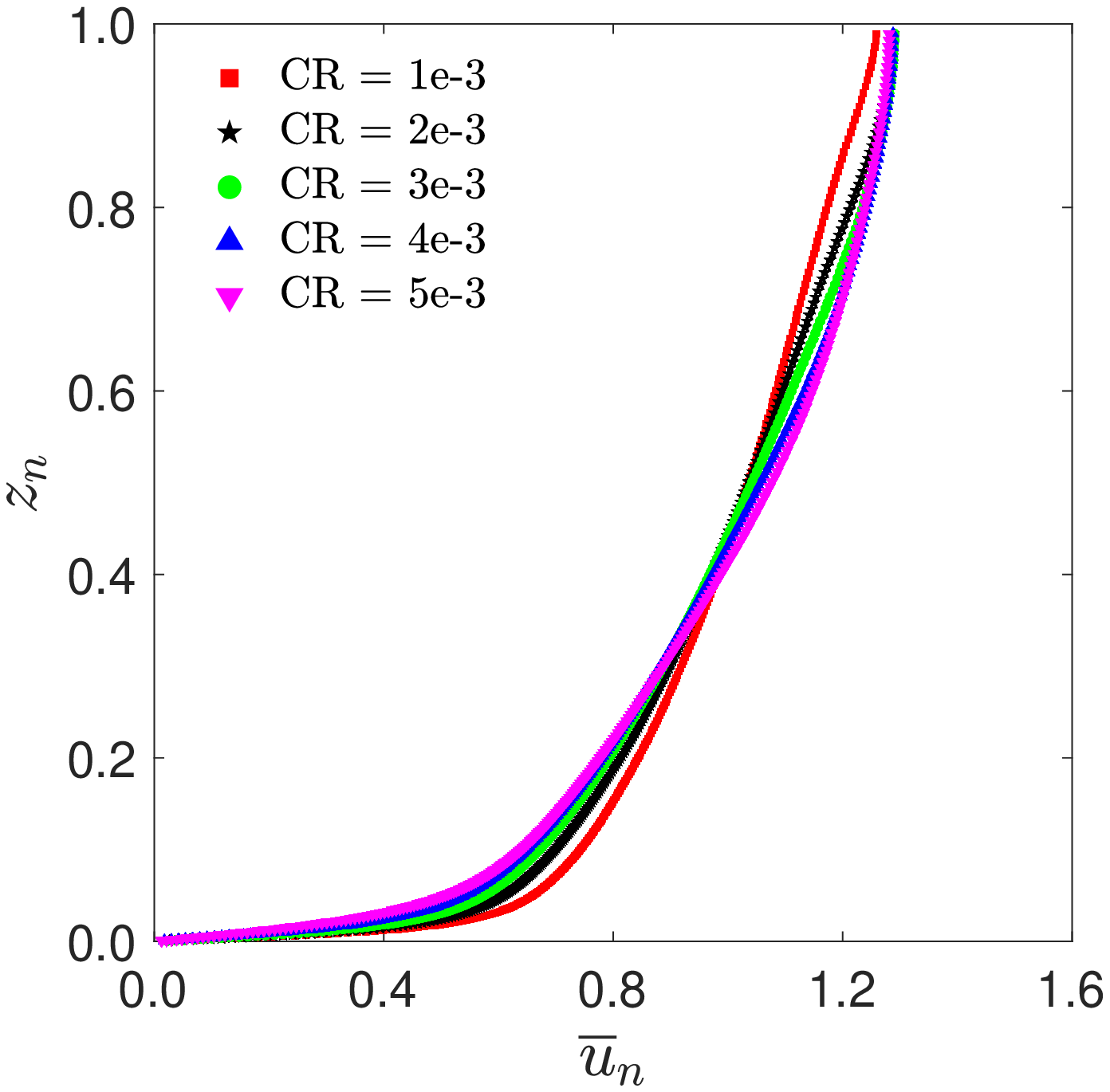}\\
  \includegraphics[height=2.2in]{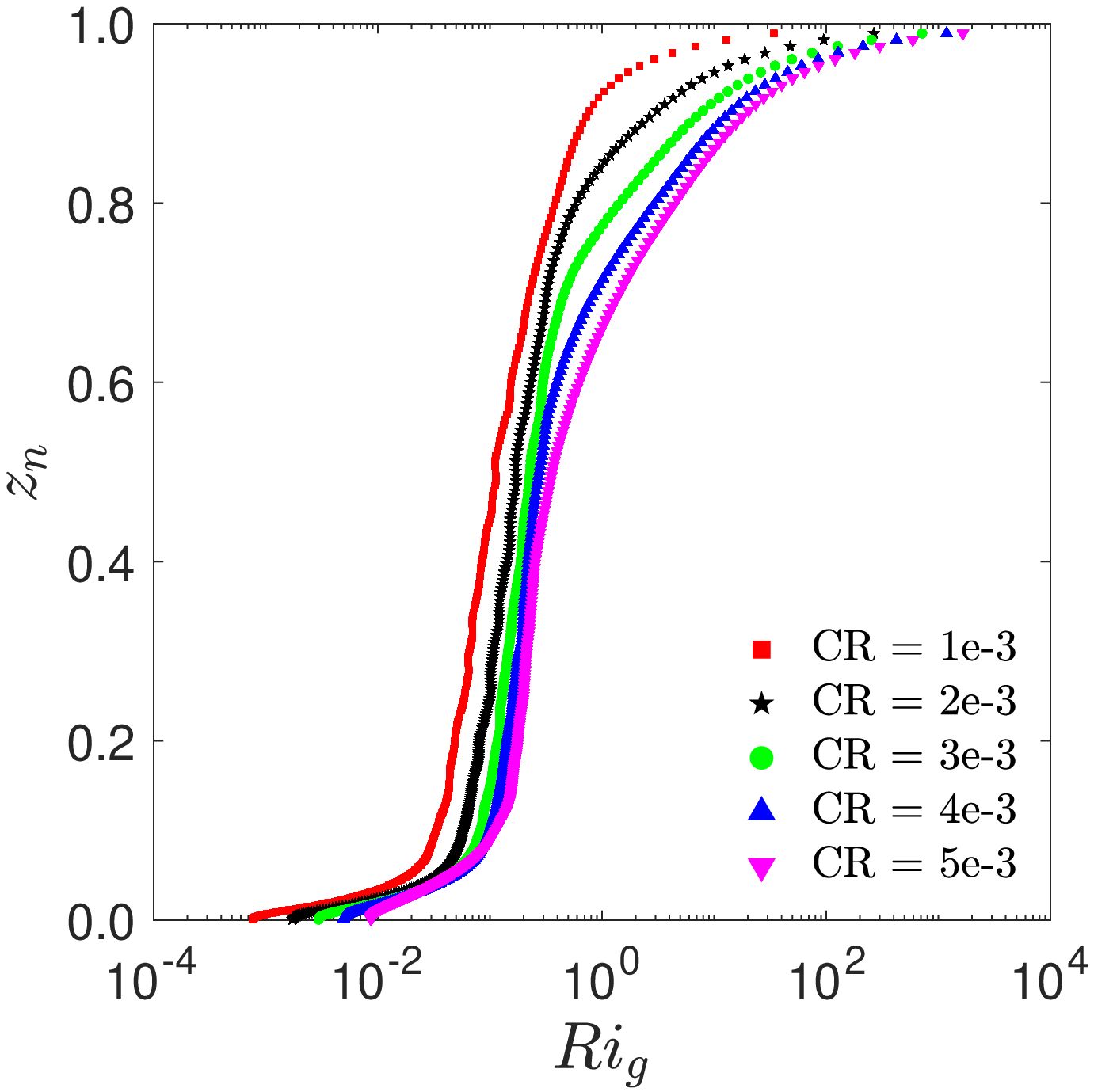}
  \hspace{0.3in}
  \includegraphics[height=2.2in]{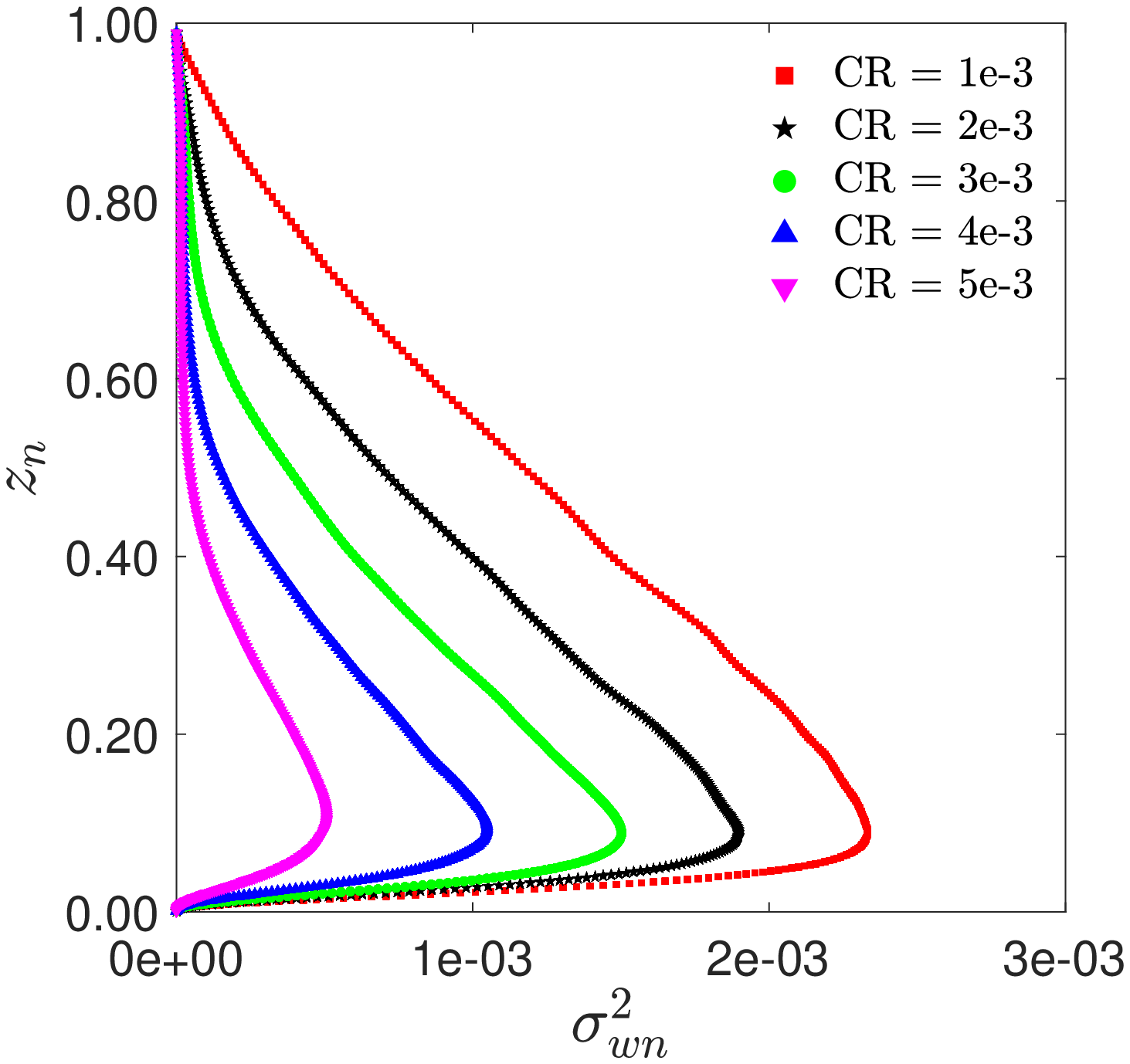}
\caption{Vertical profiles of normalized potential temperature (top-left panel), longitudinal velocity (top-right panel), gradient Richardson number (bottom-left panel), and vertical velocity variance (bottom-right panel). Simulated data from five different DNS runs are represented by different colored symbols in these plots. In the legends, $CR$ represents normalized cooling rates. All the profiles correspond to $T_n = 100$.}
\label{fig8}      
\end{figure*}

In Fig.~\ref{fig8}, vertical profiles of several key variables are plotted. It is clear that stability monotonically increases with height. As a result, turbulence in the upper part of the domain becomes quasi-laminar (especially for the runs with higher cooling rates).

For continuously turbulent stable boundary layers (SBLs), it has been frequently observed that $Ri_g$ stays below 0.2 within the SBL \citep[e.g.,][]{garratt82,nieuwstadt84,basu06}. Above the SBL, in the free atmosphere, $Ri_g$ becomes much larger. Similar behavior is noticeable in Fig.~\ref{fig8} (bottom-left panel).

\begin{figure*}[ht]
\centering
  \includegraphics[height=2.2in]{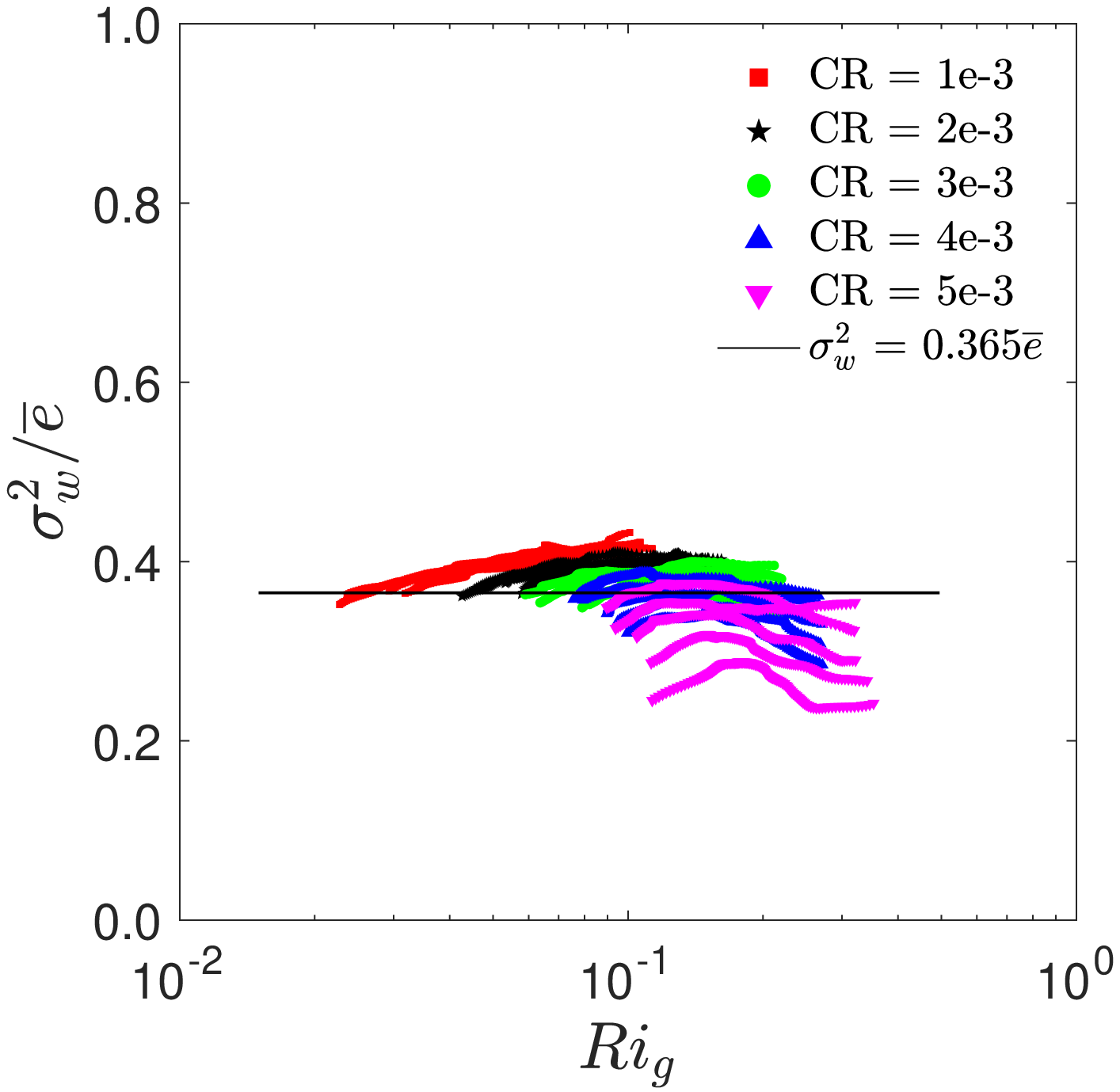}
  \hspace{0.3in}
  \includegraphics[height=2.2in]{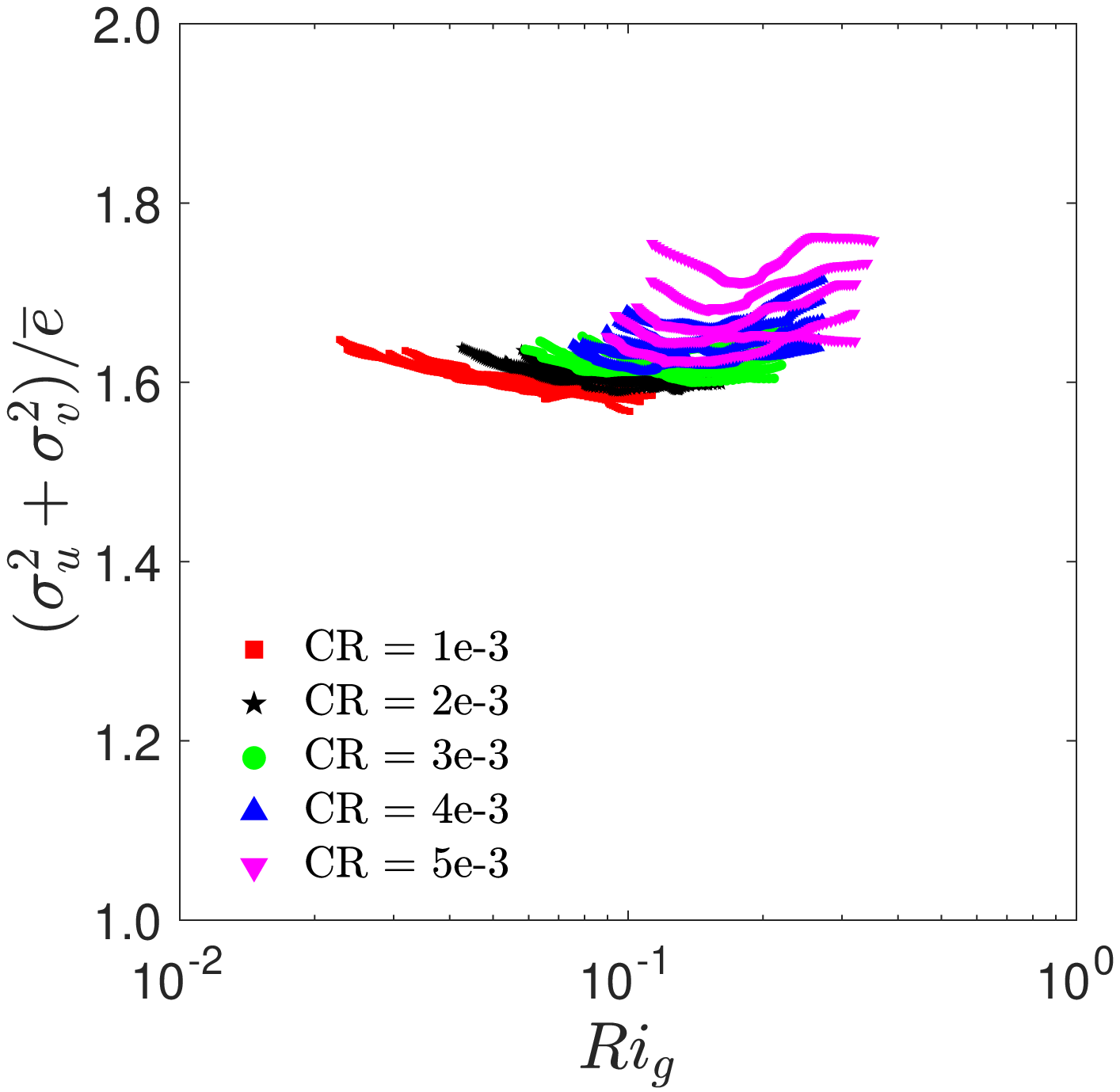}
\caption{Dependence of vertical and horizontal velocity variances on gradient Richardson number. The variances are normalized by TKE. Simulated data from five different DNS runs are represented by different colored symbols in these plots. In the legends, $CR$ represents normalized cooling rates.}
\label{fig9}      
\end{figure*}

We would like to point out that our DNS results are also in agreement with the celebrated `local scaling' hypothesis by \cite{nieuwstadt84}. By utilizing the observational data from the Cabauw tower, \cite{nieuwstadt84} showed that normalized variances remain more or less constant for a wide range of stability conditions. \cite{basu06} analyzed datasets from field campaigns, wind tunnel, and large-eddy simulations and confirmed the original findings of Nieuwstadt. In Fig.~\ref{fig9}, normalized variances from our DNS runs are shown.  

Recently, Basu et al.~\cite{basu20} found that for $0 < Ri_g < 0.2$, $\overline{\varepsilon} = 0.23 \overline{e} S$ and $\overline{\varepsilon} = 0.63 \sigma_w^2 S$. Thus, one can easily deduce that $\sigma_w^2/\overline{e}  = 0.365$. This relationship is overlaid on the DNS data in the left panel of Fig.~\ref{fig9}. Except for the data from the simulation with the highest cooling rate, this relationship is reasonably valid. Based on the LES data, Basu et al.~(2006) reported: $\sigma_w^2/\overline{e} = 0.39$. Our DNS-based result is remarkably close to this previous finding. 

\begin{figure*}[ht]
\centering
  \includegraphics[height=2.2in]{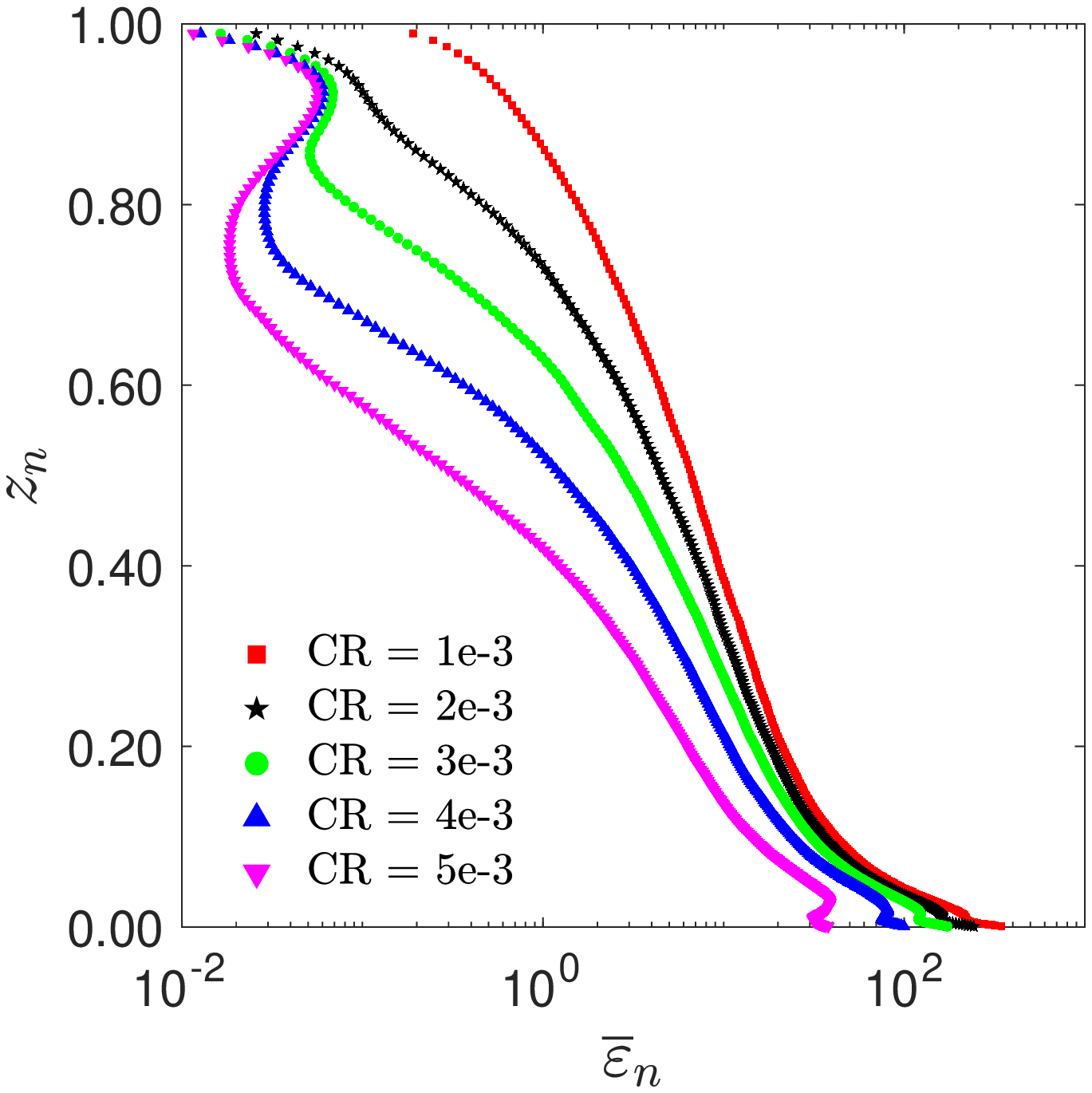}
  \hspace{0.3in}
  \includegraphics[height=2.2in]{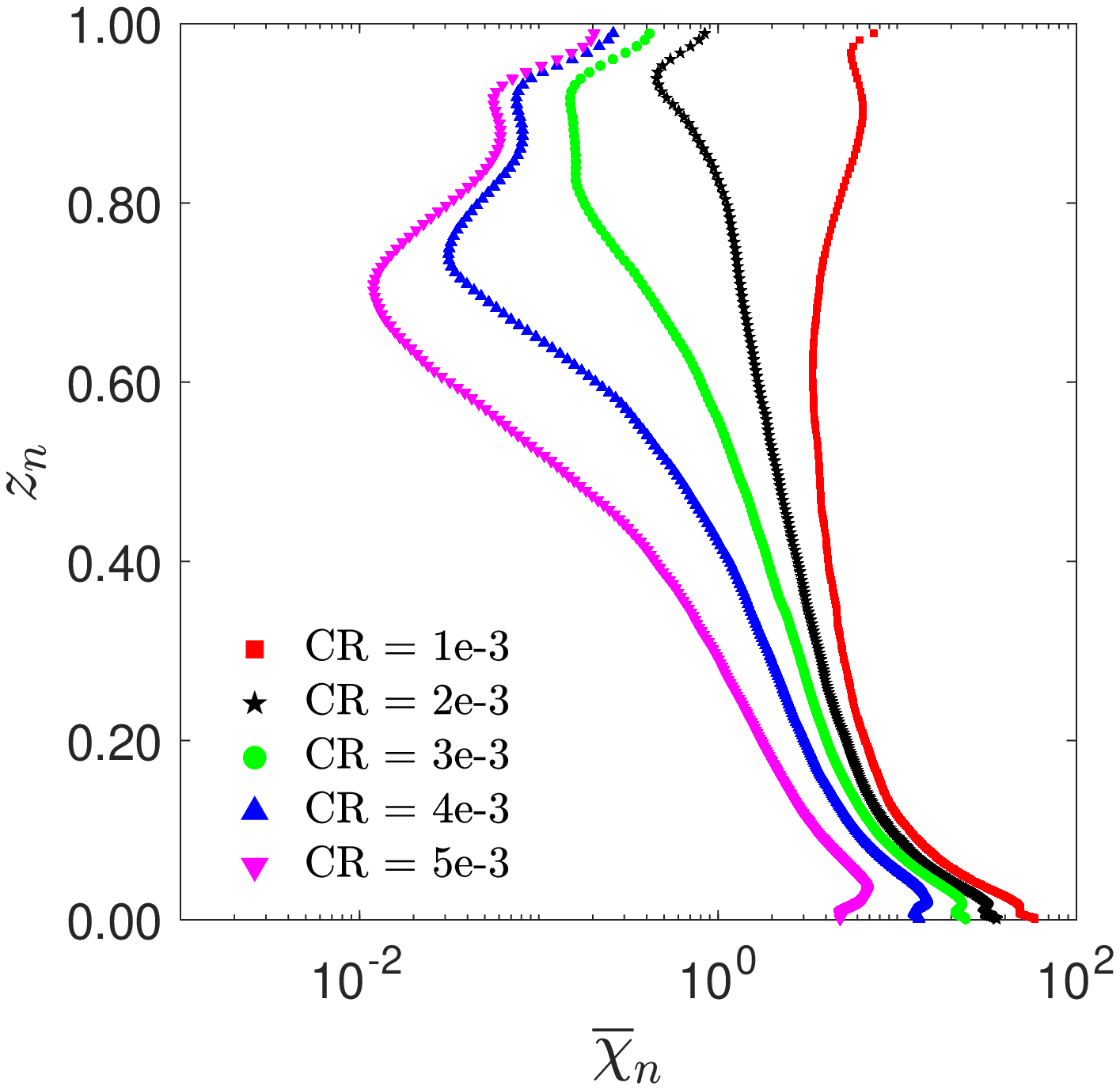}\\
  \includegraphics[height=2.2in]{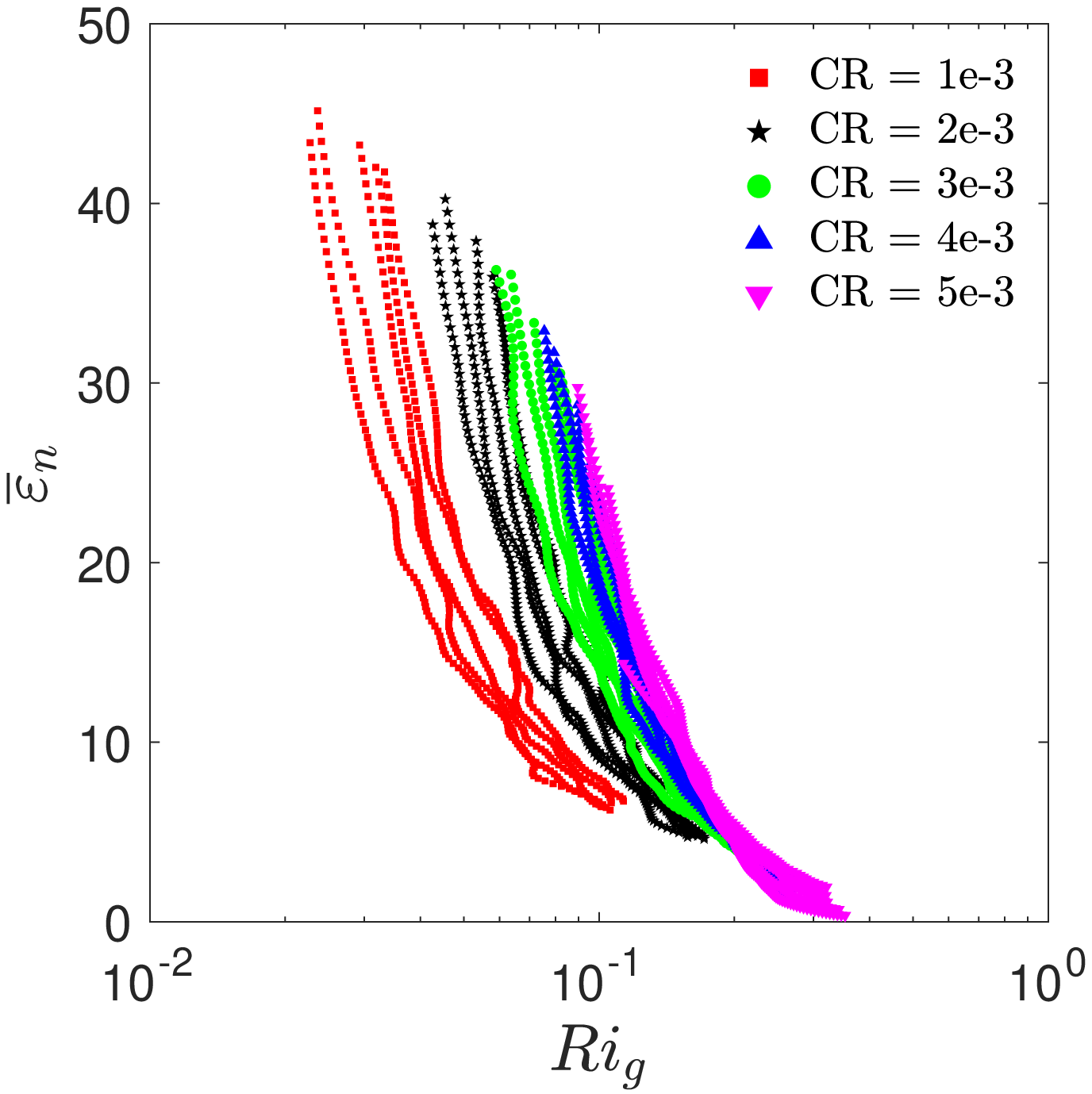}
  \hspace{0.3in}
  \includegraphics[height=2.2in]{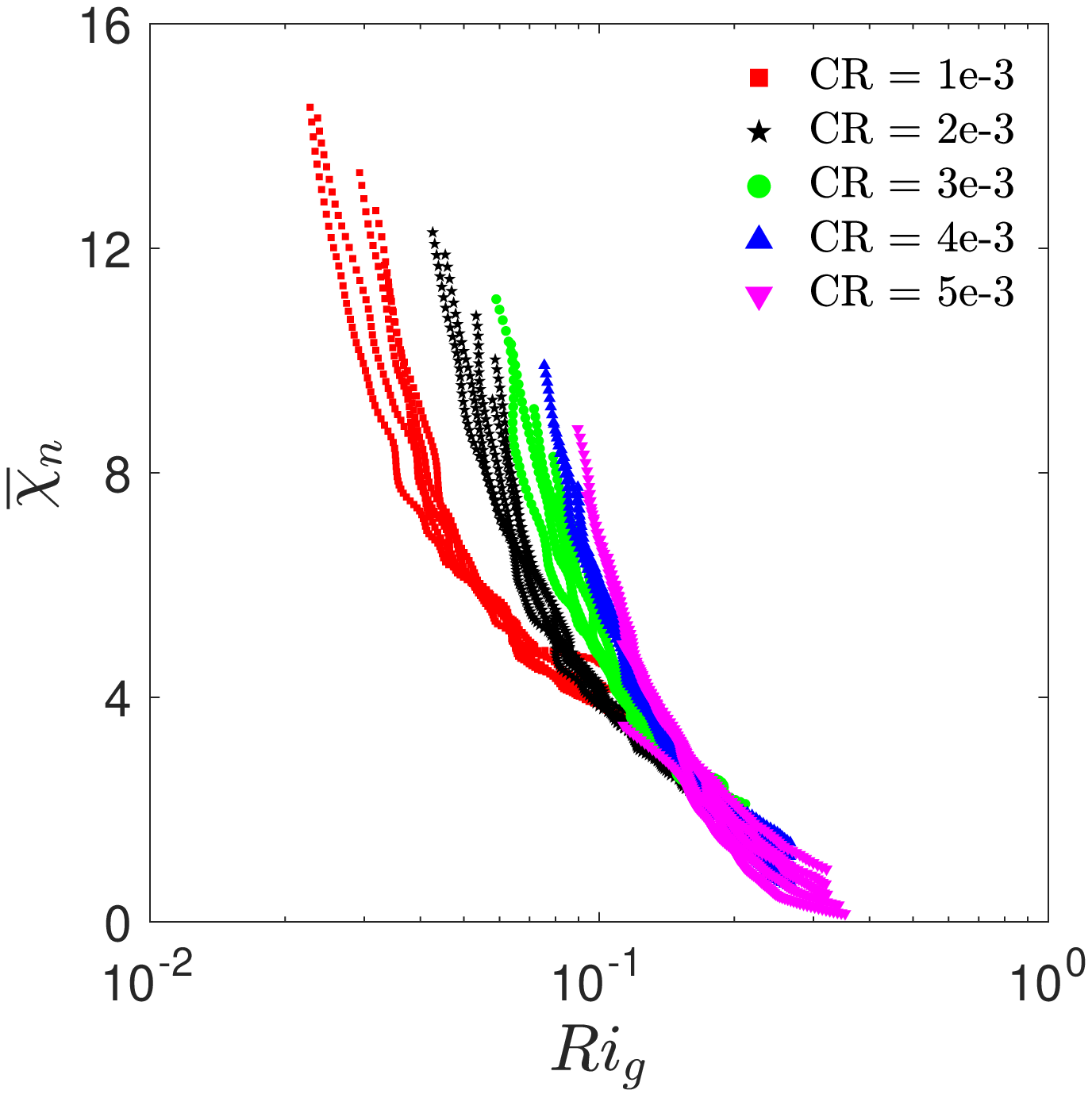}
\caption{Vertical profiles of normalized energy dissipation rate (top-left panel) and normalized dissipation of temperature variance (top right panel). All the profiles correspond to $T_n = 100$. Bottom panel: dependence of the dissipation rates on gradient Richardson number. Simulated data from five different DNS runs are represented by different colored symbols in these plots. In the legends, $CR$ represents normalized cooling rates. }
\label{fig10}      
\end{figure*}

The vertical profiles of dissipation rates are shown in the top panel of Fig.~\ref{fig10}. As expected, the dissipation rates decrease with increasing height. For $z/h < 0.1$, due to the viscous effects, the values of the dissipation rates are very high. Thus, for the computations of various length scales, we disregarded data from this region.

In our DNS runs, the bulk Reynolds number ($Re_b$) is fixed along with the initial turbulence level. Given this setup, the production of shear-generated turbulence remains the same across all the simulations. However, the destruction of turbulence due to the buoyancy effects is more predominant for the runs with higher imposed cooling rates. As a result, under steady-state condition, the dissipation rates are lower in the simulations with higher cooling rates. Such behavior can be clearly seen in both the top and bottom panels of Fig.~\ref{fig10}.

\bibliographystyle{spbasic_updated}      % basic style, author-year citations
\bibliography{CHI}   % name your BibTeX data base

\end{document}